\documentclass[preprint]{aastex}

\usepackage{lscape}
\usepackage{epsfig}
\usepackage{graphicx}
\usepackage{latexsym}
\usepackage{amssymb}
\usepackage{amsmath}
\usepackage{longtable}

\bibstyle{apj}

\newcommand{\cha}{[3.6]}
\newcommand{\chb}{[4.5]}
\newcommand{\chc}{[5.8]}
\newcommand{\chd}{[7.8]}

\def\Mpc{~$\mathrm{Mpc}$}
\def\mag{~$\mathrm{mag}$}
\def\asc{~$\mathrm{arcsec}$}

\def\mum{~$\mathrm{\mu m}$}

\def\othree{\log([O\textsc{iii}]\lambda~5007/H\beta)}
\def\ntwo{\log([N\textsc{ii}]\lambda~6584/H\alpha)}
\def\dex{$\mathrm{dex}$}
\def\mujy{$\mathrm{\mu Jy}$}

\shorttitle{The Aromatic Features in Very Faint Dwarf Galaxies}
\shortauthors{Wu et al.}

\begin{document}
\sloppy

\title{The Aromatic Features in Very Faint Dwarf Galaxies}
\author{Ronin Wu\altaffilmark{1} David W. Hogg\altaffilmark{1,2} and John Moustakas\altaffilmark{1,3}}
\altaffiltext{1}{Center for Cosmology and Particle Physics, Department of Physics, New York University, 4 Washington Place, New York, NY 10003 USA}
\altaffiltext{2}{Max-Planck-Institut f\"ur Astronomie, K\"onigstuhl 17, D-69117, Heidelberg, Germany}
\altaffiltext{3}{Center for Astrophysics and Space Sciences, University of California, San Diego, 9500 Gilman Drive, La Jolla, CA 92093-0424 USA}

\begin{abstract}
\par{We present optical and mid-infrared photometry of a statistically complete sample of 29 very faint dwarf galaxies~($M_{r}>-15$\mag) selected from the SDSS spectroscopic sample and observed in the mid-infrared with \textsl{Spitzer} IRAC. This sample contains nearby~(redshift$~z\lesssim 0.005$) galaxies three magnitudes fainter than previously studied samples. We compare our sample with other star-forming galaxies that have been observed with both IRAC and SDSS. We examine the relationship of the infrared color, sensitive to PAH abundance, with star-formation rates, gas-phase metallicities and radiation hardness, all estimated from optical emission lines.}
\par{Consistent with studies of more luminous dwarfs, we find that the very faint dwarf galaxies show much weaker PAH emission than more luminous galaxies with similar specific star-formation rates. Unlike more luminous galaxies, we find that the very faint dwarf galaxies show no significant dependence at all of PAH emission on star-formation rate, metallicity, or radiation hardness, despite the fact that the sample spans a significant range in all of these quantities. When the very faint dwarfs in our sample are compared with more luminous~($M_{r}\sim -18$\mag) dwarfs, we find that PAH emission depends on metallicity and radiation hardness. These two parameters are correlated; we look at the PAH--metallicity relation at fixed radiation hardness and the PAH--hardness relation at fixed metallicity.  This test shows that the PAH emission in dwarf galaxies depends most directly on metallicity.}
\end{abstract}


\section{Introduction}

\par{Star formation, dust and ultraviolet~(UV) and infrared~(IR) radiation in galaxies are all closely inter-related. Interstellar dust grains, excited by UV photons emitted by young stars, re-emit photons in the IR. The IR luminosity of a galaxy is found to be directly proportional to its star-formation rate~(SFR), but the relationship between star-formation and IR luminosity is complicated by uncertainty about dust~\citep{kennicutt-araa-1998}.}

\par{The spectra of star-forming galaxies in the mid-infrared~(MIR, $\lambda\sim$~3-40\mum) are generally dominated by five prominent features at 3.3, 6.2, 7.7, 8.6 and 11.3\mum. These features are now commonly attributed to small~(5\AA) polycyclic aromatic hydrocarbons~(PAH) grains~\citep{leger-apj-1984,allenmandola-apjs-1989,tielens-esasp-1999} in optical-vibrational modes excited by UV and optical photons emitted by B stars~\citep{uchida-apj-1998,li-apj-2002,peeters-apj-2004,draine-apj-2007a}. If we can understand the formation, destruction and UV illumination of PAH molecules in the interstellar medium~(ISM), these features can provide a tracer of star-formation in galaxies without highly uncertain assumptions about dust.}

\par{There has been quite a bit of work on PAH as tracers of SFR in galaxies~\citep{peeters-apj-2004,calzetti-apj-2005,wu-apj-2005}. However, PAH contents in low-luminosity galaxies appear to be exceptional. Interest in the PAH emission of low-luminosity systems was motivated by observations of NGC~5253 and \textsc{II}~Zw~40~\citep{roche-mnras-1991}. The PAH features are absent from their spectra despite large star-formation rates and substantial dust contents. A PAH deficit is also found in the MIR spectrum of SBS~0335-052, one of the most metal-poor galaxies known~\citep{izotov-apj-1997}. Despite very strong star formation activity, PAH features were absent from its MIR spectrum~\citep{thuan-apj-1999,dale-aj-2001,houck-apjs-2004}. Why are these star-forming dwarf galaxies different from star-forming galaxies of higher luminosity? The launch of \textsl{Spitzer} Space Telescope~\citep{werner-apjs-2004}, with its unprecedented sensitivity and angular resolution, has made it possible for us to make observations of large number of nearby dwarf galaxies. \textsl{Spitzer} studies on individual star-forming dwarf galaxies generally show very weak PAH features in their spectra~\citep{wu-apj-2006,jackson-apj-2006}.}

\par{Comparing normal galaxies with $M_{r} < -18$\mag~in MIR and optical colors, it is clear that visually bluer~(more star-forming) objects have stronger PAH emission, and visually redder galaxies have weaker PAH emission~\citep{forster-aa-2004, pahre-apjs-2004, hogg-apj-2005}. However, dwarf galaxies mostly appear blue and yet generally have weaker than expected PAH emission when compared to normal star-forming galaxies of similar optical color. This trend has been found in several studies with galaxies selected from \textsl{Spitzer} and optical surveys~\citep{hogg-apj-2005,rosenberg-apj-2006,wuh-apj-2007}. Are PAH molecules truly absent from dwarf galaxies, or does the PAH abundance decrease as luminosity or mass decreases? What are the causes of this difference in dwarf galaxies? Are PAH molecules destroyed by strong star-forming radiation fields? Is PAH creation or excitation inefficient in these systems because of chemical abundances or for other reasons? Is the geometrical distribution of PAH molecules is very different in dwarf galaxies? Or is it due to some other radiative transfer mechanisms that we have not yet understood?}

\par{One of the proposals is that the PAH molecules are destroyed in the harsh starburst radiation fields. These star-forming dwarf galaxies are often at a stage of massive starbursts. The hard UV radiation emitted by the massive young stars in the low-metallicity regions is capable of destroying small dust grains and PAH~\citep{plante-apj-2002}. It has been found that the MIR flux to the Far-IR flux ratio of late-type dwarf galaxies decreases as the UV flux increases~\citep{boselli-aa-1998}. The spectroscopic study of H\textsc{ii} regions in M101 also shows that the equivalent width of the 7.7\mum~PAH feature is well correlated with the ionization index, [Ne\textsc{iii}]/[Ne\textsc{ii}]~\citep{gordon-apj-2008}. Another related source of destruction could be supernova~(SN) driven shocks. SN explosions occur more frequently per unit stellar mass in these low-luminosity systems~\citep{mas-hesse-aa-1999}. Using [Ne\textsc{iii}]/[Ne\textsc{ii}] and [Fe\textsc{ii}]/[Ne\textsc{ii}] ratios to trace UV radiation hardness and SN contents, \citet{ohalloran-apj-2006} show that the SN driven shocks are likely to play some role in PAH destruction.}

\par{Another proposed cause is that low-metallicity, specially a lack of carbon, in dwarf galaxies prevents the synthesis of PAH molecules. PAH synthesis in the ISM may occur through mass-loss winds from asymptotic giant branch~(AGB) carbon-rich stars~\citep{latter-apj-1991}. If these blue dwarf galaxies are truly young, there may not be enough time to produce sufficient carbon for PAH synthesis in these systems~\citep{engelbracht-apj-2005}. This explanation maybe be applicable to the low-metallicity dwarf, SBS 0335-052; the spectrum of SBS 0335-052 appears to be dominated by extremely young star-forming episodes~\citep{dale-aj-2001, houck-apjs-2004,engelbracht-apj-2008}. However, a study of 15 local group dwarf galaxies shows evidence against this explanation~\citep{jackson-apj-2006}. All these PAH-decifient dwarf galaxies have formed the bulk of their stars more than 2 Gyr ago. It is unlikely that the shortage of AGB stars in dwarf galaxies should be the cause of PAH deficit in these dwarfs.}

\par{In this work, we select 29 extremely low luminosity~($M_{r}>-15$\mag) galaxies from the Sloan Digital Sky Survey~(SDSS,~\citealt{york-aj-2000}) to study the PAH content in these extremely faint dwarf galaxies. This sample is complete and includes by far the lowest luminosity galaxies studied statistically in the MIR. We describe the data in \S 2. We present our results in \S 3, and discuss them in \S 4. Throughout, we adopt a standard $\Lambda$CDM world model with $\Omega_{\Lambda}=0.7$, $\Omega_{m}=0.3$ and $H_{0}=72~\mathrm{km~s^{-1}~Mpc^{-1}}$. Magnitudes are given in the AB system.}

\section{Data}

\par{In our analysis, the 29 very faint galaxies are compared to normal galaxies~($M_{r}<-19$\mag) and dwarf galaxies of intermediate luminosity~($-19<M_{r}<-17.5$\mag) to see how the MIR emission varies with luminosity. The comparison galaxies are chosen from available public surveys and all have been observed with both \textsl{Spitzer} IRAC and SDSS. There are in total 1166 comparison galaxies. In this section, we describe our data selection and reduction.}

\subsection{Sample Selection}

\par{Our primary sample comprises 29 dwarf galaxies of very low luminosity~($M_{r}>-15$\mag, $\sim L^{*}/100$). The galaxies in our sample are selected from the SDSS DR2~\citep{abazajian-aj-2004} spectroscopic survey. There are 28,089 galaxies in the low-redshift~($0.0033<z<0.05$) catalog of the New York University Value-Added Galaxy Catalog~(NYU-VAGC;~\citealt{blanton-apj-2005a}). We select from this subset all galaxies that have $r<16$\mag~and $M_{r}>-15$\mag, and are within a recession velocity of $2000~\mathrm{km~s^{-1}}$~($\sim 28$\Mpc). Selection with the above criteria yields 29 very low luminosity galaxies. The procedure of obtaining the photometric and spectroscopic data for this complete sample is described in the following subsections, and the results are presented in Table(\ref{gals}).}
\par{The $M_{r}$ values used in this work are quoted from the New York University Value-Added Galaxy Catalog~(NYU-VAGC). In the catalog, the distance to each individual galaxy is calculated with the Local Group barycentric redshifts and a local velocity field model established by \citet{willick-apj-1997}. All $M_{r}$ values are extinction-corrected using the Galactic dust map based on the 100\mum~emission~\citep{schlegel-apj-1998} and are k-corrected using the specialized software~\citep{blanton-aj-2003}.}

\subsection{IRAC imaging}

\par{Our targets were observed with 150s exposures and a 5-point Gaussian dither using all four channels~(3.6\mum, 4.5\mum, 5.8\mum~and 7.8\mum) of IRAC as a part of \textsl{Spitzer} GO-2 program~(program NO.: 20210, PI: David W. Hogg). The Basic Calibrated Data~(BCD) produced by each exposure was generated by automated \textsl{Spitzer} processing pipeline version 13.2.}
\par{The first part of the pipeline is a combination of the data reduction modules, which are responsible of dark current subtraction, bad pixels removal~(including muxbleed and radhit corrections), array response linearization and flat-fielding. The data uncertainty for each pixel is estimated as the Poisson error in the number of the detected electrons and the readout noise added in quadrature. Calibration of BCD images is carried out by using a set of calibrator stars to obtain flux conversion factors. Correction of color due to zodiacal light and correction of distortion due to the IRAC array offset from the optical axis of \textsl{Spitzer} were also performed as part of the calibration process~\citep{reach-pasp-2005}. The sky dark frame is generated by observing a pre-selected low zodiacal background in the north ecliptic cap and subtracted from the BCD data in the pipeline.}
\par{All BCD images of each galaxy in each channel were co-added with pointing refinement to produce post-BCD~(PBCD) images at \textsl{Spitzer} Science Center~(SSC). PBCD-mosaic images are generated by averaging the BCD images. The uncertainty of coordinates given in each BCD image~($\sim 0.5$\asc) was improved by using the positions of the Two Micron All Sky Survey~(2MASS, \citet{struskie-aj-2006}) sources to within $0.3$\asc~for the PBCD-mosaic images. Hot or dead pixels are recorded in the mask files and rejected during coaddition accordingly. Additional detail regarding BCD and PBCD-mosaic image production can be found in {\em IRAC Data Handbook}.}

\subsection{SDSS imaging}
\par{The SDSS-mosaic images and SDSS-mosaic error images of our targets are part of the Fifth Data Release~(DR5,~\citealt{adelmen-mccarthy-apjs-2007}). Our SDSS-mosaic images are generated from the SDSS reconstructed frames. The images are smoothed with an $0.8$\asc~Gaussian kernel. This additional smoothing on top of the SDSS seeing~($\sim1.4$\asc) ensures that the SDSS-mosaic images have a point-spread function~(PSF) similar to IRAC images whose PSF has a mean full width half maximum~(FWHM) of $\sim 1.7$\asc. The sky level of each mosaic image is estimated and subtracted using median-smoothed sky vector of the immediately adjacent fields during mosaicing. The SDSS-mosaic error image of each frame records the total noise contributed by dark current, read noise, and Poisson noise from photon statistics. Mosaics in all five bands of SDSS~($u$, $g$, $r$, $i$ and $z$) for each galaxy in our sample were generated.}

\subsection{Photometry}
\par{Our photometry measurements are based on the PBCD-mosaic images and the PBCD-uncertainty images provided by SSC. The IRAC photometric data presented in Table(\ref{gals}) is based on modeling the IRAC imaging with a model constructed from the SDSS imaging of the same sources. The IRAC and SDSS apparent magnitudes for the 29 very low-luminosity galaxies are therefore custom-made "model" magnitudes. Detailed description of the model is included in Appendix A.}
\par{The first step of our photometry measurement is to apply our photometry model to the SDSS-mosaic images and compare our results with the photometry measurement made with the pure exponential model by the SDSS DR5 photometric pipeline~\citep{lupton-aj-1999}. This step serves as a sanity check of our photometric model. The detailed description of our model can be found in Appendix A. The magnitudes calculated by us differ from the magnitudes calculated with the results returned from the SDSS DR5 pipeline by $\sim 1.21$~percent on average in all five bands. This difference comes from small differences between our photometric model and the slightly cruder pure exponential model used in the pipeline. The SDSS magnitudes in our final result are Galactic extinction corrected.}
\par{We apply our model to the IRAC remapped PBCD-mosaic images to measure the flux received in the IRAC channels. For each of our target, we obtained measurements in 9 bands~(IRAC: 3.6\mum, 4.5\mum, 5.8\mum~and 7.8\mum~(hereafter, $\cha$, $\chb$, $\chc$ and $\chd$); SDSS: $u$, $g$, $r$, $i$ and $z$). The results are presented on on the AB system in Table(\ref{gals}). Galaxies included in this sample are of very low redshifts~($z\lesssim 0.005$). Therefore, we apply no K-corrections on our photometry.}

\subsection{\textsl{Spitzer} Public Data}
\par{We build a comparison sample with the galaxies found in the overlapping region of the SDSS footprints and two sets of \textsl{Spitzer} public data, the First Look Survey~(FLS,~\citealt{lacy-apjs-2005}) and the \textsl{Spitzer} Wide-area InfraRed Extragalactic survey~(SWIRE,~\citealt{lonsdale-pasp-2003},~hereafter, FLS/SDSS and SWIRE/SDSS galaxies). We match the coordinates of the galaxies in the two sets of \textsl{Spitzer} surveys with the coordinates of the galaxies in SDSS DR5. The selected galaxies have less than 1\asc~of difference between their \textsl{Spitzer} and SDSS coordinates. For these galaxies, the optical photometric and spectroscopic data used in this work is taken directly from the DR5 pipeline. The redshift of the FLS/SDSS and SWIRE/SDSS galaxies ranges from $0.006$ to $0.359$, which is not a big span. No K-correction has been applied to the photometric data.}
\par{It was impractical to apply the photometric modeling to all of the comparison data. The IRAC photometry of the FLS/SDSS galaxies was measured from the \textsl{Spitzer} BCD images through $9.2$\asc~diameter apertures with rejection of images containing bad pixels or cosmic rays within the aperture, according to the \textsl{Spitzer}-provided mask files. Background was determined by taking a median in an annulus of inner radius 18\asc~and outer radius 28\asc.  The photometry of images produced by multiple dithers was averaged together~\citep{hogg-apj-2005}. There are 240 FLS/SDSS galaxies in the comparison sample.}
\par{The SWIRE DR2 of the northern sky, Elaic\_n1, Elaic\_n2 and Lockman Hole, has around 18 $\mathrm{deg^{2}}$ of the area overlapping with footprints of the SDSS~\citep{davoodi-mnras-2006}. The IRAC photometry for SWIRE/SDSS galaxies was derived from the \textsl{Spitzer} BCD images in \textsl{Spitzer} Science Center using the S11.4.0 pipeline, in which the sources were extracted using the SExtractor software~\citep{bertin-aaps-1996}, and the aperture fluxes were extracted within five separate apertures, $1.4$, $1.9$, $2.9$, $4.1$ and $5.8$\asc~\citep{lonsdale-pasp-2003}. The apparent magnitudes of the SWIRE/SDSS galaxies in this work are calculated with the $2.9$\asc~aperture fluxes. There are 926 SWIRE/SDSS galaxies in the comparison sample.}
\par{Fig.(\ref{color_absmr}) shows the relationship between the $[g-r]$ and $M_{r}$ at $z=0$. Comparing to the FLS/SDSS and SWIRE/SDSS galaxies in Fig.(\ref{color_absmr}), the mean $M_{r}$ of the very low luminosity galaxies in our sample stands out by 3\mag~fainter than the mean $M_{r}$ of the dwarf galaxies~($M_{r}>-19$\mag) that are found among the FLS/SDSS and SWIRE/SDSS galaxies.}

\subsection{Spectroscopy}
\par{The SDSS spectra are taken with 3\asc~optical fibers plugged into the spectrograph focal plane. The spectra cover a wavelength range of 3800\AA~to 9200\AA~with a resolution of $\lambda/\Delta\lambda\sim2000$~\citep{york-aj-2000}. The spectra are calibrated with observations of F subdwarfs through the pipeline~\citep{abazajian-aj-2004}.}
\par{The rest-frame emission line flux for all galaxies is measured with the IDL code, \emph{platefit}, which is developed primarily by C. Tremonti and is specially designed to fit a stellar continuum to the SDSS spectra so to recover weak nebular features. Detailed discussion of \emph{platefit} can be found in \citet{tremonti-apj-2004}, and a discussion of the adaptation of the code to non-SDSS spectra can be found in \citet{lamareille-aa-2006}. We briefly summarize the procedure of the line measurements in this section.}
\par{The code adopted a stellar population synthesis model~\citep{bruzual-mnras-2003} which incorporates an empirical spectral library well matched to the SDSS spectra and produces model template spectra. After subtracting the continuum from the SDSS spectra by fitting a combination of a few model template spectra and removing the remaining residuals, all the emission lines are fitted with Gaussians simultaneously. In this fitting process, all the Balmer lines are required to have the same line-width and velocity offsets, and all the forbidden lines are required to have the same line-width and velocity offsets. The ratio of the $[N\textsc{ii}]\lambda~6548$ and $[N\textsc{ii}]\lambda~6584$ lines is set to be equal to the theoretical value 3. The result of our line measurements is presented in Table(\ref{gals}).}

\section{Results}

\subsection{Infrared Color}
\par{The emission measured by the fourth channel~(CH4, 7.8\mum) of IRAC is dominated by the 7.7\mum~small-PAH~($<10^{3}$ C atoms) feature and continuum from hot dust. The emission measured by the first channel~(CH1, 3.6\mum) of IRAC is dominated by starlight; CH1 observations are expected to trace the underlying stellar mass of the galaxy. It has been suggested that IRAC CH1 can be used as a direct tracer of the stellar mass free of dust obscuration~\citep{pahre-apjs-2004}. Although the PAH feature at 3.3\mum~can contribute to the CH1 flux, the spectra of the 17 \textsl{Spitzer} Infrared Nearby Galaxies Survey~(SINGS, \citealt{kennicutt-pasp-2003}) galaxies fitted by a dust model which consists of specified mixtures of carbonaceous grains and amorphous silicate grains~\citep{draine-apj-2007a} shows that the IRAC CH1 and CH2 fluxes are generally dominated by the stellar emission~\citep{draine-apj-2007b}.}
\par{We examine the effect by comparing $\chb-\chd$ color with $\cha-\chb$ color in Fig.(\ref{color_12_42}) with galaxies located at z$<$0.05. The three stars indicate the blackbody spectra of three different stellar surface temperatures~(3,000$\mathrm{K}$, 5,000$\mathrm{K}$, 9000$\mathrm{K}$ red, yellow and blue, respectively). The colors of the blackbody spectra are computed with k\_project\_filters.pro contained in the kcorrect software v4\_1\_4~\citep{blanton-aj-2003}. In general, the scatter of the $\cha-\chb$ color can be reasonably well-explained by the range of the blackbody colors, but the $\chb-\chd$ color for the galaxies is much redder than the pure blackbodies. This is consistent with the idea that the $\cha-\chb$ color is dominated by the stellar color, while the $\chb-\chd$ color depends more on the ISM contents. Therefore, we construct the $\cha-\chd$ color as a measure of the small-PAH emission plus the dust continuum normalized by the stellar mass.}

 \subsection{Optical Color}
\par{Fig.(\ref{color_gr_14}) shows the relationship between the $\cha-\chd$ and $[g-r]$ colors for star forming galaxies. Normal galaxies with $M_{r}<-19$\mag~generally lie on a trend with visually red~(old) galaxies showing blue $\cha-\chd$ and visually blue~(young) galaxies showing red $\cha-\chd$. At the visually bluer end~($[g-r]\lesssim 0.6$\mag) of Fig.(\ref{color_gr_14}), the $\cha-\chd$ color becomes bluer as $r$ band luminosity~($M_{r}$, indicated by color, see Fig.(\ref{color_absmr}) caption) decreases while the $[g-r]$ color remains the same~($0.2\lesssim [g-r] \lesssim 0.7$\mag)~(see also \citealt{hogg-apj-2005,rosenberg-apj-2006}). Comparing dwarf galaxies of intermediate luminosity~($-19<M_{r}<-17.5$\mag) with dwarf galaxies of very low luminosity~($M_{r}>-17.5$\mag), we find that the median $[g-r]$ colors~(0.36 and 0.33\mag, respectively) of these two groups of galaxies are similar, however, median $\cha-\chd$ values~(0.77 and $-$0.44, respectively) decrease substantially with the stellar mass. The dot-dashed line in Fig.(\ref{color_gr_14}) at $\cha-\chd=-1.58$ indicates the prediction for a completely dust-free environment where the color is estimated with stellar population model Starburst99~\citep{leitherer-apjs-1999}, extrapolated from the 3.6\mum~emission~\citep{helou-apjs-2004}. It is interesting to note that the dwarf galaxies, even the very faint ones, have $\cha-\chd$ about one mag redder than the dust-free case.}

\subsection{Emission-line Classification}
\par{Comparing 28 SINGS H\textsc{ii} nuclei galaxies with 24 SINGS low luminosity AGNs, it was noted that the H\textsc{ii} nuclei, in general, have a smaller fraction of their dust masses contributed by small PAHs than AGN galaxies~\citep{draine-apj-2007b}. The possibility that the AGN sources can directly excite PAH emission complicates any observational relation between PAH emission and star-formation~\citep{smith-apj-2007}. In our M$_{r}>-15$\mag~sample we do not expect such low-mass galaxies to host AGNs, so when comparing to galaxies of higher luminosity, we would like to keep only star-forming galaxies in the comparison sample.}
\par{The selection of star-forming galaxies in the comparison sample is performed with a Baldwin-Phillips-Terlevich diagram (\citealt{baldwin-pasp-1981}, hereafter BPT diagram, see also: \citealt{veilleux-apjs-1987,kewley-apj-2001}) shown in Fig.(\ref{o3_n2}). We use the empirical demarcation curve given by \citet{kauffmann-mnras-2003} to define the star-forming galaxies. The selected galaxies all have $S/N>3$ in H$\alpha$, H$\beta$ and the two nebular lines, [O\textsc{iii}]$\lambda$5007 and [N\textsc{ii}]$\lambda$6584. By this selection, of the 829 galaxies plotted in Fig.(\ref{o3_n2}), 442 are classified as star-forming galaxies. Six galaxies in the very low-luminosity sample are excluded from this figure. The spectrum of SDSS\_J115825.59$+$505501.4 has S/N too low to yield reasonable line measurements. The spectrum of SDSS\_J115132.93$-$022222.0 has no data at $6200\AA<\lambda<7000\AA$. SDSS\_J124157.06$+$034909.3, SDSS\_J124433.42$+$014412.9, SDSS\_J132818.63$+$673800.3 and SDSS\_J135723.58$+$053425.0 appears old; they do not show any emission lines in their spectra. There are six galaxies with S/N$<$3 for two of the nebular lines. We plot these in Fig.(\ref{o3_n2}) as limits. The twenty-three very low-luminosity galaxies plotted in Fig.(\ref{o3_n2}) are all classified as star-forming galaxies according to the demarcation.}
\par{With this comparison sample of 442 star-forming galaxies, we examine the relationship between the $\cha-\chd$ and three optical spectra-based properties, the specific star-formation rate~(SFR), the oxygen abundance, and the radiation hardness, in the rest of this section.}

\subsection{EW(H$\alpha$)}
\par{In Fig.(\ref{pah_haew}), we compare the $\cha-\chd$ color with EW(H$\alpha$), which effectively measures the ratio of massive ionizing stars to low mass red giants and is used as an indicator of the star-formation rate per unit luminosity or mass~\citep{kennicutt-apj-1994}. To eliminate the uncertainty of the result caused by comparing two quantities measured with different apertures~(the SDSS fiber aperture is 3\asc), we require that all galaxies plotted in Fig.(\ref{pah_haew}) should have the $[g-r]$ color should have no greater than a 0.2\mag~difference between the fiber-measured and photometric colors. We divide the galaxies into three groups according to their $M_{r}$ values. The linear regression of $\cha-\chd$ color as a function of $\log$(EW(H$\alpha$)) gives the slopes of 1.4, 1.0 and 0.1 for the most luminous~($-23.0<M_{r}<-19.0$\mag), the intermediate~($-19.0<M_{r}<-17.5$\mag) and the least luminous~($M_{r}>-17.5$\mag) groups, respectively. Fig.(\ref{pah_haew}) shows that the dependence of the $\cha-\chd$ color on the specific SFR becomes weaker with decreasing luminosity of galaxies. At high luminosity, the linear relationship between $\cha-\chd$ color and EW(H$\alpha$) is:}
\begin{equation}
[3.6]-[7.8]=1.40\, \log(EW(H\alpha))-0.44
\end{equation}

\subsection{Oxygen Abundances and Radiation Hardness}
\par{A gas-phase oxygen abundance, $12+\log(O/H)$, is estimated from the optical spectrum by
\begin{equation}
12+\log(O/H)=8.73-0.32\,\othree\,\ntwo
\end{equation}
The relationship between $\cha-\chd$ color and this metallicity estimate is presented in Fig.(\ref{color_metal}). The metallicity estimates have uncertainties greater than 0.25\dex~when $12+\log(O/H)<8.1$~\citep{pettini-mnras-2004}.}
\par{The median $12+\log(O/H)$ for the seventeen very low luminosity galaxies with good line measurements is 7.66~\dex. For galaxies with oxygen abundance greater than 8.1, the median $\cha-\chd$ color is 1.4~\mag~with a standard deviation of 0.58~\mag. In the low-oxygen-abundance group ($12+\log(O/H)\leq 8.1$), the median $\cha-\chd$ color is 0.19~\mag~with a standard deviation of 0.66~\mag.}
\par{To examine the relationship between $\cha-\chd$ and radiation hardness, we use $\othree$~as an indicator of the ionization parameter. Fig.(\ref{pah_o3}) shows $\cha-\chd$ as a function of $\othree$. The $\cha-\chd$ value decreases as $\othree$ increases. The most luminous galaxies~($M_{r}<-19$\mag) show the reddest $\cha-\chd$ color~(1.48\mag~in median) and the softest radiation fields~($-0.40$ in $\othree$~median). Galaxies of intermediate luminosity~($-19<M_{r}\leq -17.5$\mag) have a median $\cha-\chd$ color of 0.63~\mag~and median $\othree$~of 0.06~\dex. The least luminous group~($-17.5<M_{r}\leq 13.5$\mag) show the bluest $\cha-\chd$ color~(-0.19~\mag~median) and the hardest radiation median $\othree$ of 0.27~\dex. This trend and Fig.(\ref{pah_o3}) imply that the strength of the small-PAH feature at 7.7\mum~in star-forming galaxies is generally inversely related to the hardness of galaxy radiation fields.}
\par{In general, galaxies show a relationship between metallicity and radiation hardness. This complicates the interpretation of results when treating the oxygen abundance and the radiation hardness as independent variables. To improve this from Fig.(\ref{color_metal}) and Fig.(\ref{pah_o3}), we plot in Fig.(\ref{irc_delta}) the relationships between the $\cha-\chd$ color and metallicity or radiation hardness in groups of galaxies selected by radiation hardness or metallicity.}
\par{In Fig.(\ref{irc_delta})(left), the galaxies are divided according to radiation hardness, and in Fig.(\ref{irc_delta})(right) they are divided according to metallicity. The divisions are chosen such that in both panels, the median luminosities are $M_{r}=-17.9$, $-20.0$, and $-20.9$ from top to bottom. The $\cha-\chd$ color appears to have a stronger relationship with metallicity in the harder radiation group. The relationship between the $\cha-\chd$ color and metallicity becomes weaker as the radiation field gets softer.}
\par{A straight line is fit to each group. It appears that, when looking at the lowest metallicity and the highest radiation hardness, the $\cha-\chd$ color has a better relationship with metallicity rather than with hardness.}

\section{Discussion}
\par{We have presented MIR properties of 29 very low-luminosity galaxies with $M_{r}>-15$\mag~based on \textsl{Spitzer} IRAC observations. We presented the MIR results using the $\cha-\chd$ color. Theoretical models containing mixtures of amorphous silicate and graphite grains~(including varying amount of PAH particles) show that in a high PAH abundance environment, the $\chd$ flux is dominated by the $7.7$\mum~PAH feature; even in the absence of PAH emission a small amount of dust exposed to starlight can contribute significantly to the $\chd$ flux~\citep{draine-apj-2007a}. Because the $\cha$ flux is dominated by starlight~\citep{pahre-apjs-2004}, we will refer to the $\cha-\chd$ color as ''(PAH+dust)/star'', which effectively measures the small-PAH emission~(with potential hot-dust contribution)~normalized by the stellar mass.}
\par{If small-PAH features are generally excited by absorption of UV photons emitted by massive stars~\citep{li-apj-2002,peeters-apj-2004}, red galaxies, which are lacking young massive stars, are incapable of producing strong PAH emission. On the other hand, blue galaxies, since still young and actively forming stars, have sufficient amount of massive stars that are capable of exciting these features. Fig.(\ref{color_gr_14}) confirms this: The bulk of galaxies show a strong relationship between (PAH+dust)/star and the optical color, with blue galaxies showing much more PAH emission than red  galaxies. But also in Fig.(\ref{color_gr_14}), the dwarf ($M_{r}>-19.0$) galaxies clearly deviate from this overall trend. These dwarf galaxies appear to be blue in $[g-r]$ color, which implies strong star-forming activity, but show smaller (PAH+dust)/star ratios when compared to normal galaxies of similar colors. This result has been investigated and confirmed by previous optical and MIR observations~\citep{hogg-apj-2005, rosenberg-apj-2006}. Dividing the dwarf galaxies into two luminosity groups, $-19<M_{r}<-17.5$\mag~and $M_{r}>-17.5$\mag, the (PAH+dust)/star ratio continues to decrease as the luminosity decreases. That said, even for the faintest group, the ratio is still higher than the predicted value for a completely ISM-free environment. This result implies the existence of PAH molecules or hot dust in these very low-luminosity galaxies. The mechanism behind the deficit of PAH emission in dwarf galaxies is uncertain. It may arise from a mixture of mechanisms, for example, the particular star-formation histories, metallicities and hard radiation fields found in these galaxies~\citep{ohalloran-apj-2006,wu-apj-2006,calzetti-apj-2007,engelbracht-apj-2008}. The disentanglement of these effects, however, is difficult, because these properties of dwarf galaxies are themselves closely inter-related.}
\par{The relationship between PAH emission and the star-formation is intriguing. The MIR spectra of normal star-forming galaxies are, in general, dominated by the PAH features and dust continuum. One of the candidate sources that can excite these MIR features is the UV photons emitted by the young and massive stars. However, the carriers of the 7.7\mum~PAH feature are also likely to be destroyed by the hard radiation field in the starburst regions~\citep{madden-aa-2006}. Star-formation, on the one hand, can excite the PAH features, but, on the other hand, can also destroy the feature carriers. Fig.(\ref{pah_haew}) shows that the $\cha-\chd$ color and the logarithm of the equivalent width of the H$\alpha$ line~(star-formation rate) holds a linear relation with a slope of 1.4~\dex~per~\dex~for normal star-forming galaxies. However, the relationship between the (PAH+dust)/star ratio and SFR is much weaker for dwarf galaxies of intermediate luminosity and almost gone for dwarf galaxies of very low luminosity. When the measurement is limited to only the star-forming regions with metallicity near the Solar value, the 7.7\mum~PAH feature shows an almost linear relationship with the ionizing photon rate, which can be measured with Paschen-$\alpha$ line~\citep{calzetti-apj-2007}. However, the use of the 7.7\mum~feature as a tracer for SFR is very sensitive to the metallicity, the star-formation history, and the size of the galaxy.}
\par{Although the formation process for PAH molecules is still not well understood, if the absence of small-PAH features in dwarf galaxy spectra is a problem of formation inefficiency, then a strong dependence of the $\cha-\chd$ color on metallicity is expected. Our result in Fig.(\ref{color_metal}) shows that the (PAH+dust)/star increases with increasing oxygen abundance. In general, this result agrees with the previous MIR studies of low metallicity galaxies: Observations show that galaxies with low metallicity~(less than 0.2 Solar metallicity) tend to have lower equivalent widths for emission features at 6.2, 8, and 11.2\mum, which are features from small($\sim 10\AA$) PAH grains~\citep{wu-apj-2006,engelbracht-apj-2008}. A study of the \textsl{Spitzer} IRS low-resolution spectra of 61 SINGS galaxies fitted with a theoretical dust model, have also shown that, in low metallicity~($12+\log(O/H)<8.1$) galaxies, the fraction of the dust mass contributed by small PAHs is 1\% in median~\citep{draine-apj-2007b}, compared to a more standard fraction of 3.55\% at higher metallicity.}
\par{So far it is well accepted that the PAH features are suppressed in low-luminosity galaxies, which, in general, are metal-poor. With the complete sample of very low-luminosity galaxies presented here, it is ideal if one can establish a reliable relationship between PAH feature strength and metallicity. Although we are limited by the non-negligible uncertainties on the estimates of abundances for the lower-metallicity galaxies, Fig.(\ref{color_metal}) provides evidence that the 7.7\mum~feature is substantially weaker for low-metallicity galaxies. Better abundance estimation is probably not possible without better spectra. For galaxies at such low redshifts,  the [O\textsc{ii}]$\lambda$3726 and [O\textsc{ii}]$\lambda$3729 lines are well below the minimum wavelength of SDSS spectroscopy, so it is impossible at present to estimate the oxygen abundances of these very low-luminosity galaxies with R$_{23}$~\citep{pilyugin-apj-2005}. Direct measurement of the electron temperature of the gas with the ionization correction factor~(ICF) is difficult because the [O\textsc{iii}]$\lambda$4363 line is weak~(only 7 of the 29 very low-luminosity galaxies has S/N$>$3 in this line).  To explore in more detail the dependence of the (PAH+dust)/star ratio on metallicity, it will be important to have reliable metallicity estimates for these metal-poor galaxies.}
\par{The absence of PAH feature from the low-luminosity galaxy spectra could also be an effect of any PAH destruction process. This is demonstrated by Fig.(\ref{pah_o3}), where we show that (PAH+dust)/star ratio is inversely related to the radiation hardness~(as indicated by $\othree$). A similar relationship has been found in a previous analysis of 66 starburst galaxies observed in the MIR with the \textsl{Spitzer} Space Telescope. The equivalent width of the 7.7\mum~PAH feature is inversely related to the radiation hardness index~(a combination of [Ne\textsc{iii}]/[Ne\textsc{ii}] and [S\textsc{iv}]/[S\textsc{iii}], \citealt{engelbracht-apj-2008}). IRS observations of 26 blue compact dwarf galaxies~(BCDs) showed that the equivalent width of the 11.2\mum~ and 7.7\mum~PAH features are suppressed in harder radiation fields~(as indicated by [Ne\textsc{iii}]/[Ne\textsc{ii}], \citealt{wu-apj-2006}). The ISOCAM MIR spectra of 7 dwarf galaxies show weak small PAH features~(including 6.2, 7.7, 8.6, 11.3, and 12.6\mum) and hard radiation fields~(indicated by [Ne\textsc{iii}]/[Ne\textsc{ii}]~\citealt{madden-aa-2006}). Analysis of the five PAH features~(6.2, 7.7, 8.6, 11.3 and 12.7 \mum) in the H\textsc{ii} regions of M101 shows that the PAH equivalent width relates better to the ionization parameter, \textsc{ii}$\equiv$[Ne\textsc{iii}]/[Ne\textsc{ii}], than to the oxygen abundances; the same behavior is also seen for starburst galaxies including several dwarf galaxies~\citep{gordon-apj-2008}.}
\par{Because there are lower opacities in lower-metallicity H\textsc{ii} regions, the photons escaping from the photospheres better preserve their energies; radiation field tends to be harder in these systems because the photons have not been effectively scattered~\citep{dopita-apj-2006,bresolin-apj-2007}. Among the faintest dwarf galaxies~($M_{r}>-15$\mag), there is no significant (PAH+dust)/star--metallicity or (PAH+dust)/star--radiation hardness relation~(see Fig.(\ref{color_metal}) and Fig.(\ref{pah_o3})), even though these galaxies span $\sim1.5$~\dex~in both parameters. However, when the very faint galaxies are compared with more luminous dwarfs, we find that the PAH emission in dwarf galaxies depends more directly on metallicity. The left panel of Fig.(\ref{irc_delta}) compares the relationship between the (PAH+dust)/star and the $12+\log(O/H)$ among galaxies of similar radiation hardness. The dependence of the (PAH+dust)/star on $12+\log(O/H)$ becomes stronger when the radiation in galaxies gets harder. The right panel compares the (PAH+dust)/star with the radiation hardness of galaxies at similar oxygen abundances. The dependence of (PAH+dust)/star on radiation hardness is stronger for galaxies of high metallicity, but the (PAH+dust)/star is almost independent of radiation hardness~for metal-poor~($12+\log(O/H)<8.0$) galaxies. We do not confidently draw the conclusion that the metallicity plays a more important role in the absence of PAH for dwarf galaxies for two reasons. First, the values of $12+\log(O/H)$ in the top plot of Fig.(\ref{irc_delta})(left) have more than 0.25~\dex~of uncertainty~\citep{pettini-mnras-2004}. Second, the estimation of $12+\log(O/H)$ itself depends on the $\othree$~ratio. To further disentangle the relationships of the (PAH+dust)/star ratio with metallicity and radiation fields for these very low-luminosity galaxies, it will be necessary to have independent measurements of the oxygen abundance and the ionization parameter.} 

\acknowledgments
It is our pleasure to thank Michael R. Blanton and Christy A. Tremonti for contributions of code and data. We would also like to thank Eric F. Bell for helpful discussion and Morad Masjedi for his help with the original \textsl{Spitzer} observing proposal. We would like to thank the Sloan Digital Sky Survey and the Spitzer Space Center for making their data available. This research has made use of NASA's Astrophysics Data System Bibliographic Services and the idlutils package. RW was supported by the Spitzer GO-2 LOWLUM program~(ID: 20210, PI: David W. Hogg) and the \textsl{Spitzer} GO-5 S5 program ~(ID: 50568 and 50569, PI: David Schiminovich). DWH was partially supported by NASA (grant NNX08AJ48G) and a Research Fellowship of the Alexander von Humboldt Foundation. JM was supported by NASA (grant 06-GALEX06-0030) and \textsl{Spitzer} (grant G05-AR-50443).

\clearpage
\begin{figure}
\centering
\begin{tabular}{c}
\epsfig{file=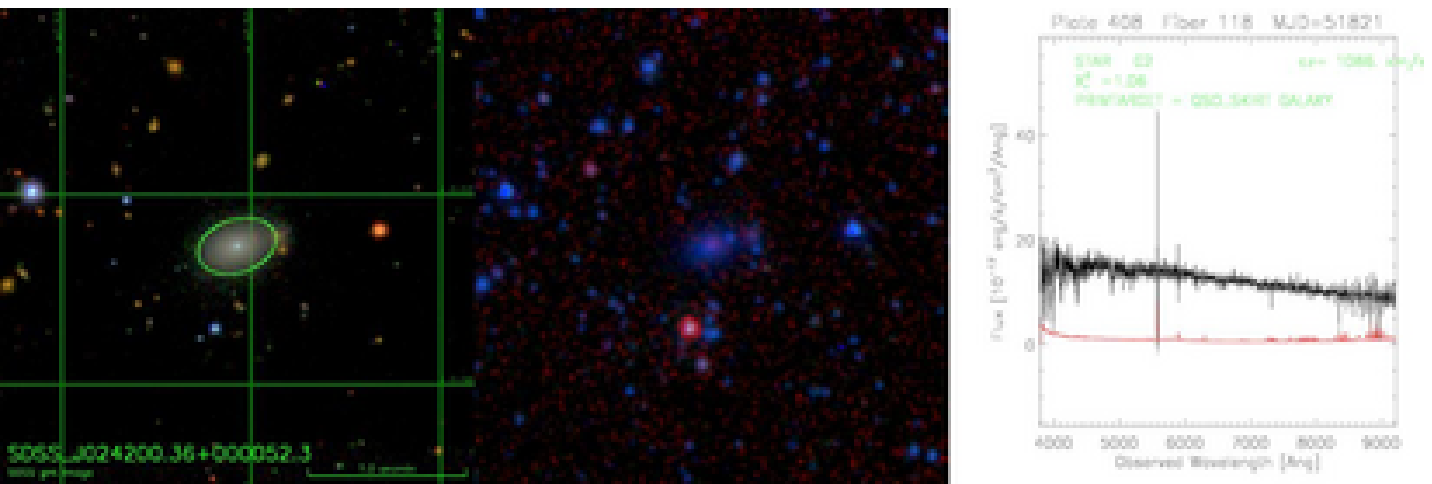,width=0.8\linewidth} \\
\epsfig{file=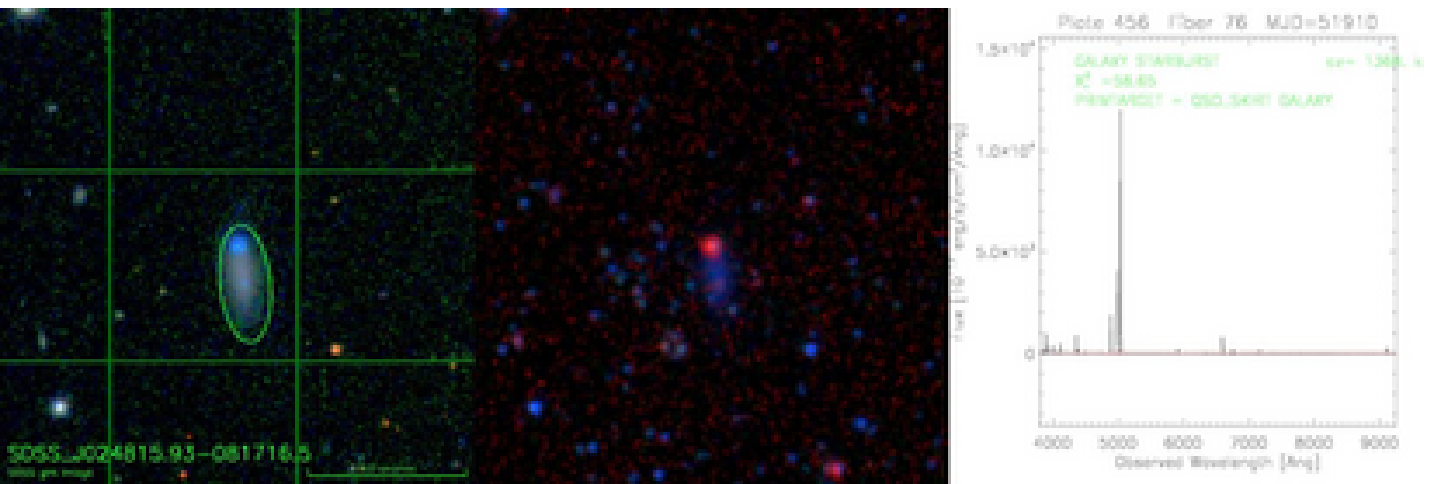,width=0.8\linewidth} \\
\epsfig{file=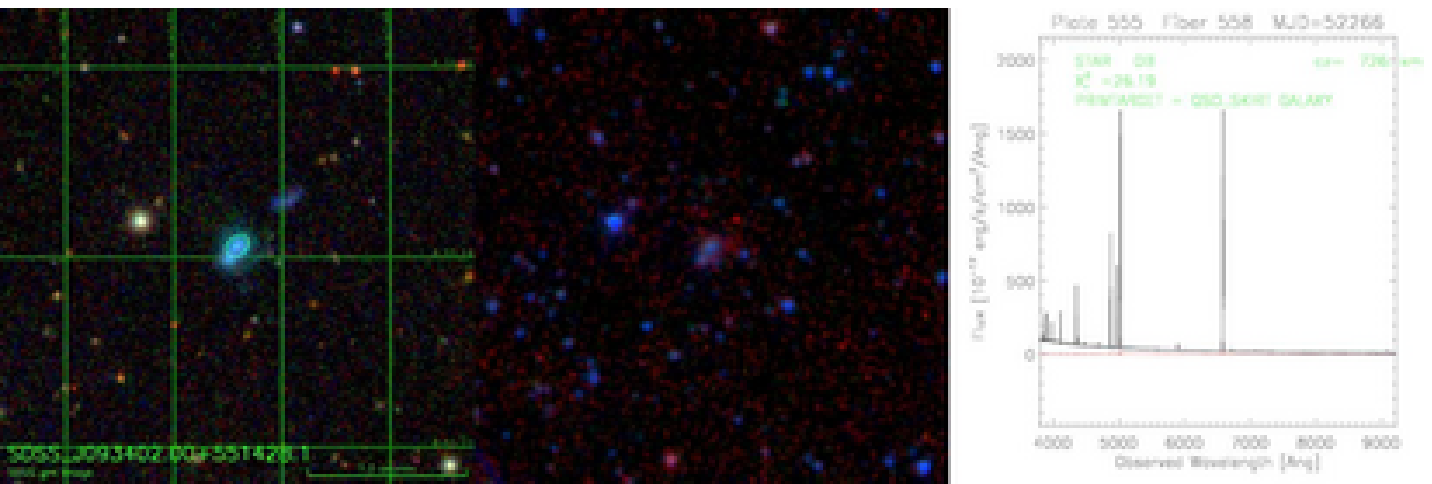,width=0.8\linewidth} \\
\epsfig{file=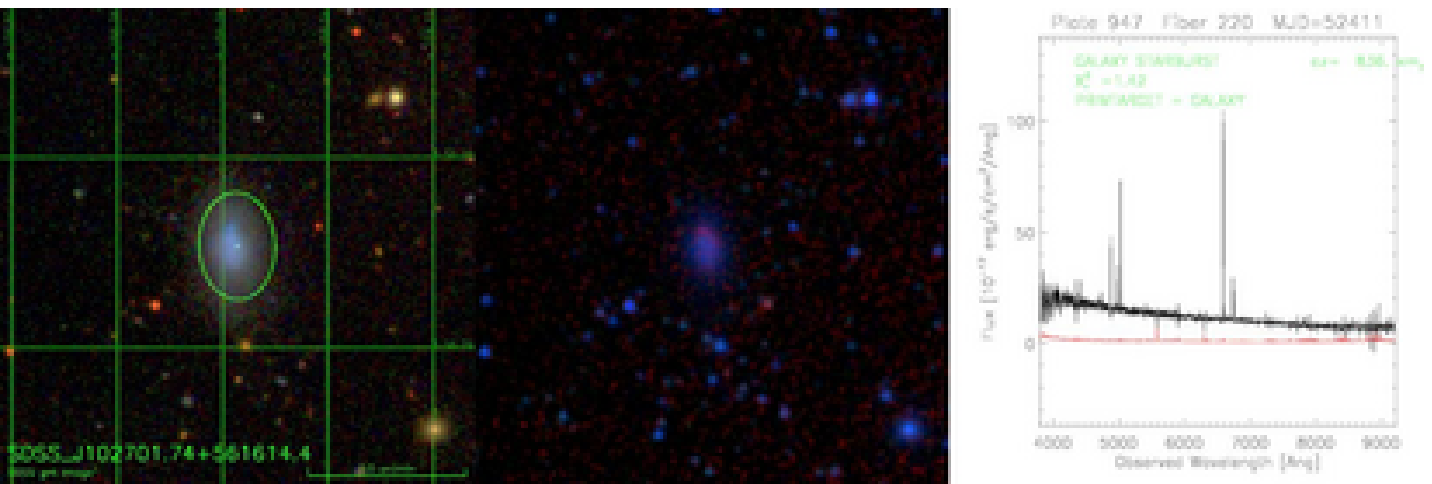,width=0.8\linewidth} \\
\end{tabular}
\figcaption{The 29 very low-luminosity galaxies studied in this work. For each galaxy, the SDSS image, the IRAC image and the SDSS spectrum are presented from left to right. The SDSS color image is made with $i$, $r$ and $g$ band images as red, green and blue colors respectively. The IRAC color image is made with $\chd$, $\chb$, and $\cha$ images as red, green and blue colors respectively.\label{galpic_0}}
\end{figure}
\addtocounter{figure}{-1}
\thispagestyle{empty}
\begin{figure}
\centering
\begin{tabular}{c}
\epsfig{file=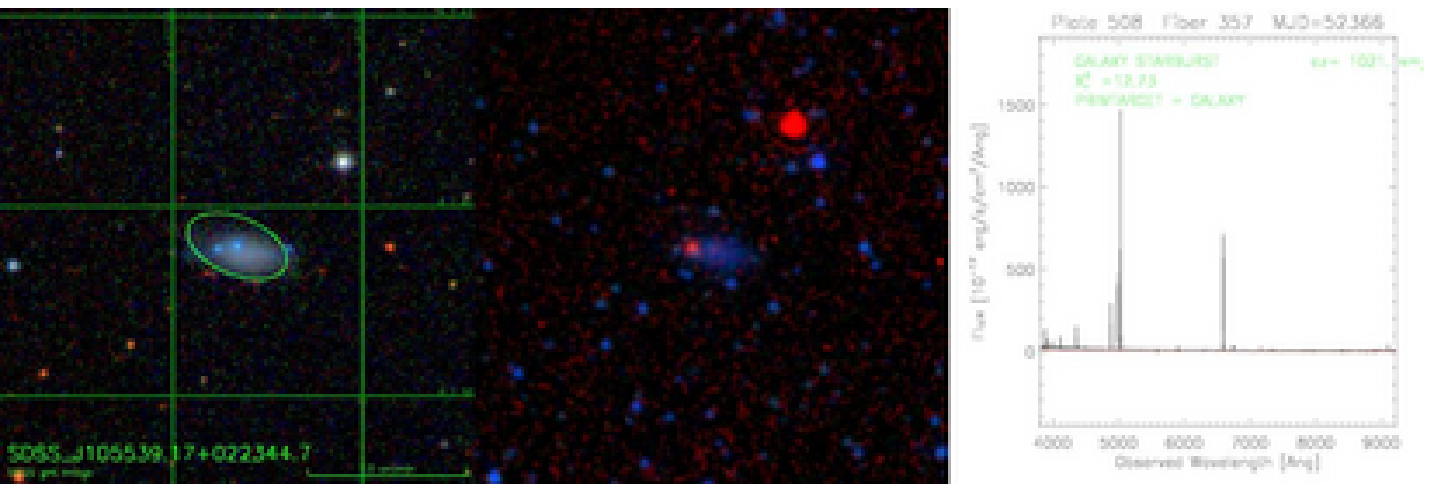,width=0.8\linewidth} \\
\epsfig{file=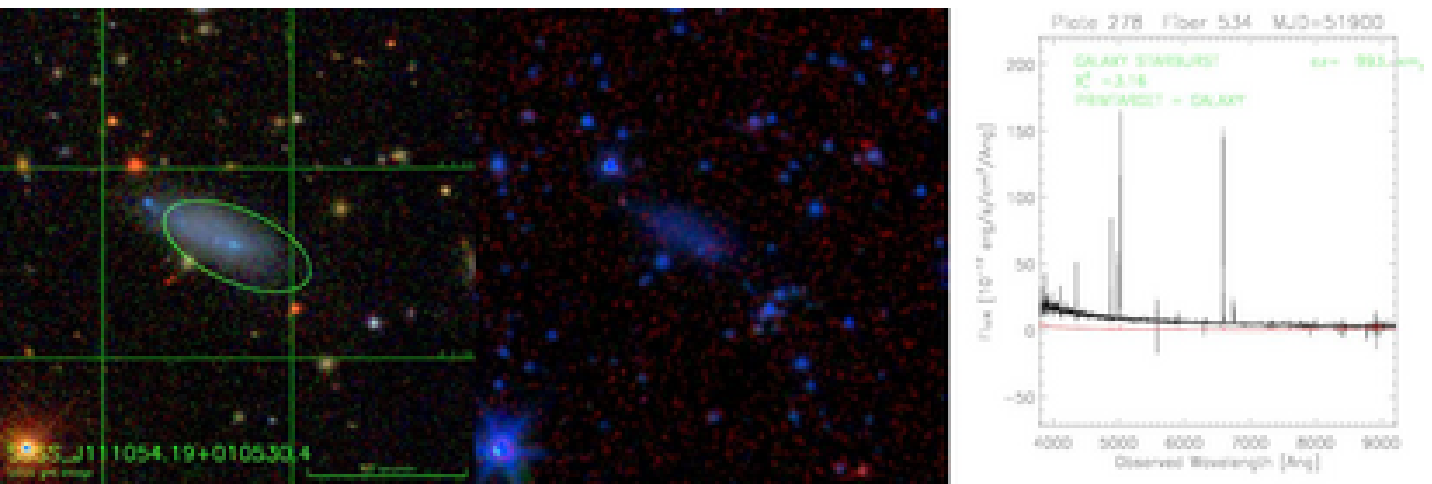,width=0.8\linewidth} \\
\epsfig{file=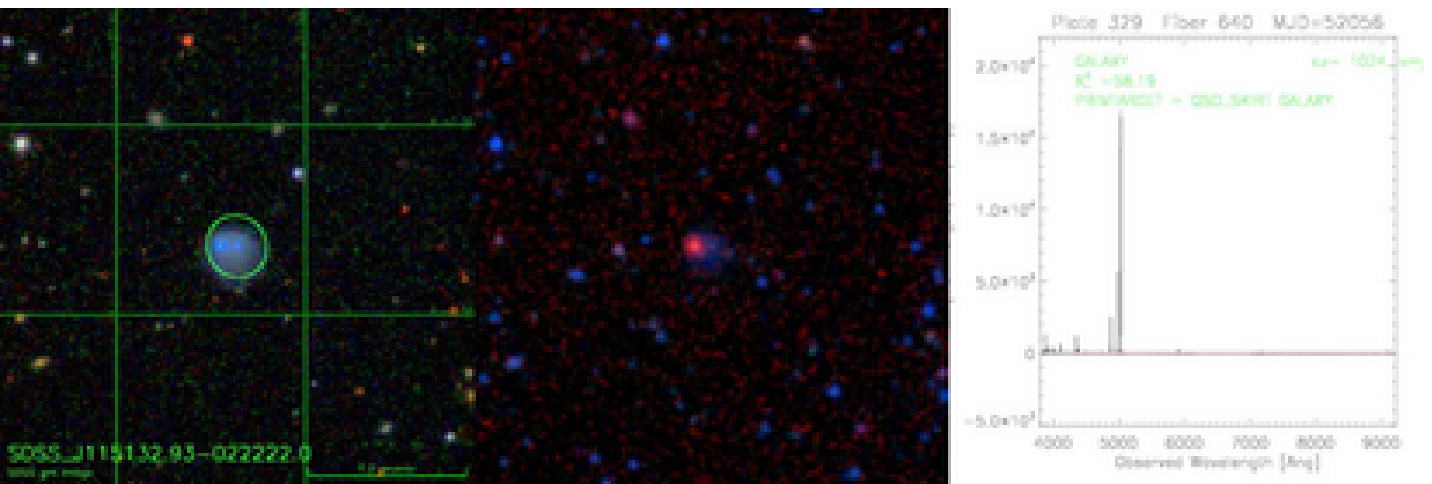,width=0.8\linewidth} \\
\epsfig{file=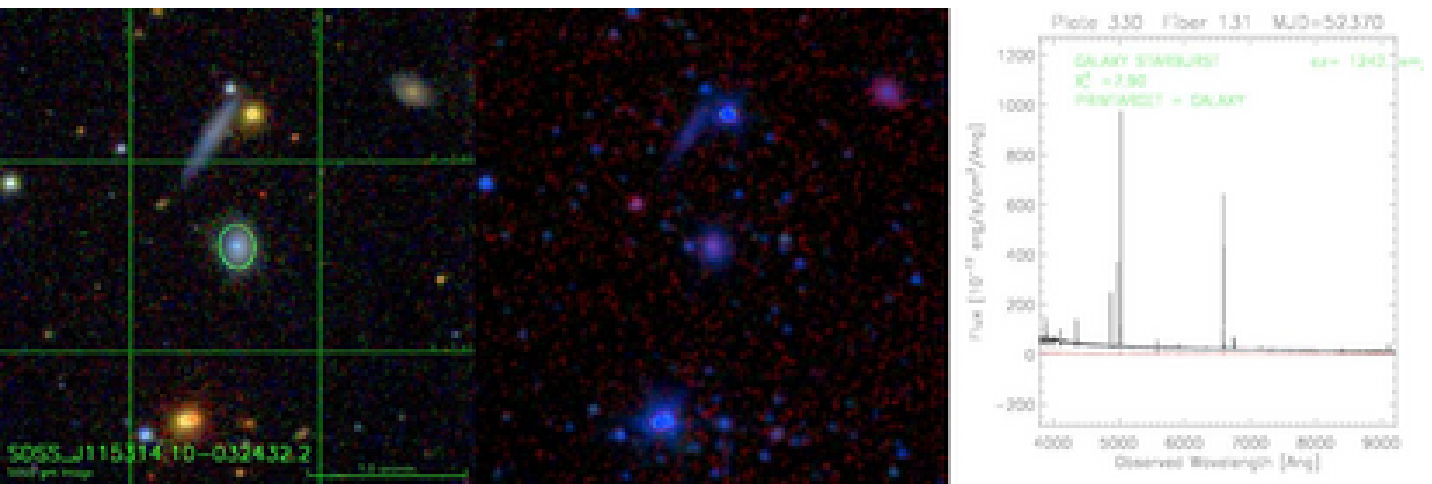,width=0.8\linewidth} \\
\epsfig{file=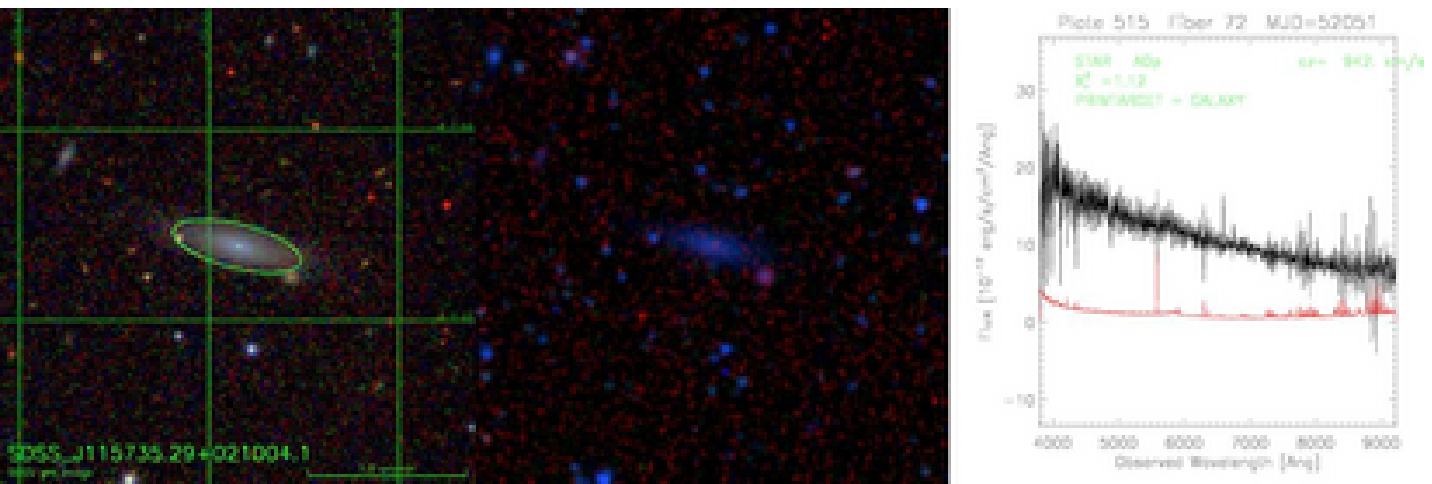,width=0.8\linewidth} \\
\end{tabular}
\figcaption{Cont.\label{galpic_1}}
\end{figure}
\addtocounter{figure}{-1}
\thispagestyle{empty}
\begin{figure}
\centering
\begin{tabular}{c}
\epsfig{file=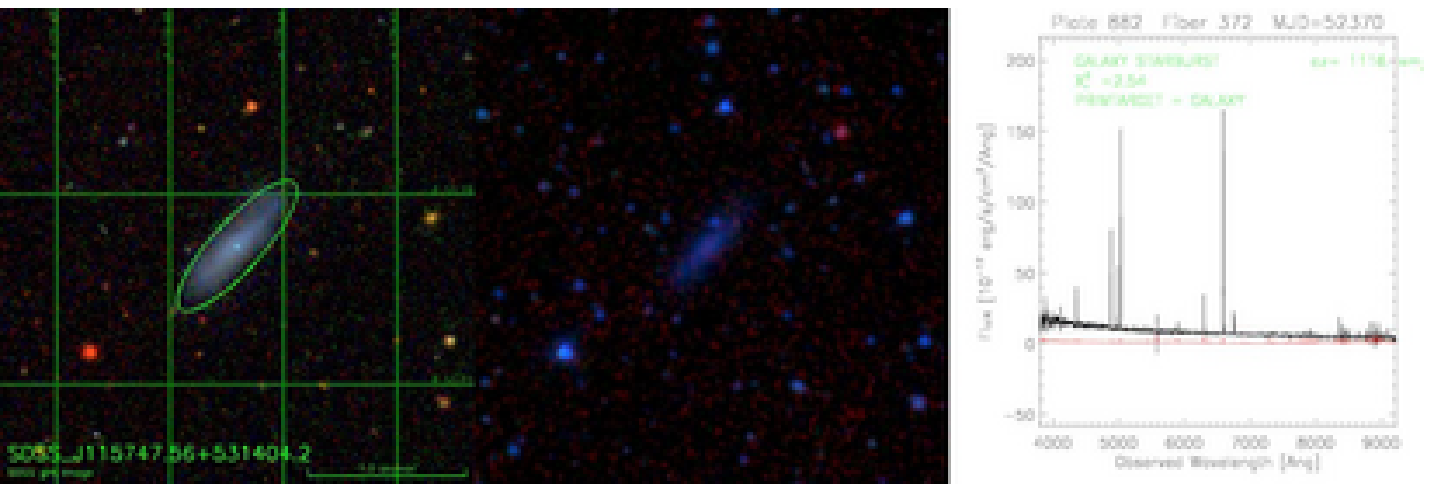,width=0.8\linewidth} \\
\epsfig{file=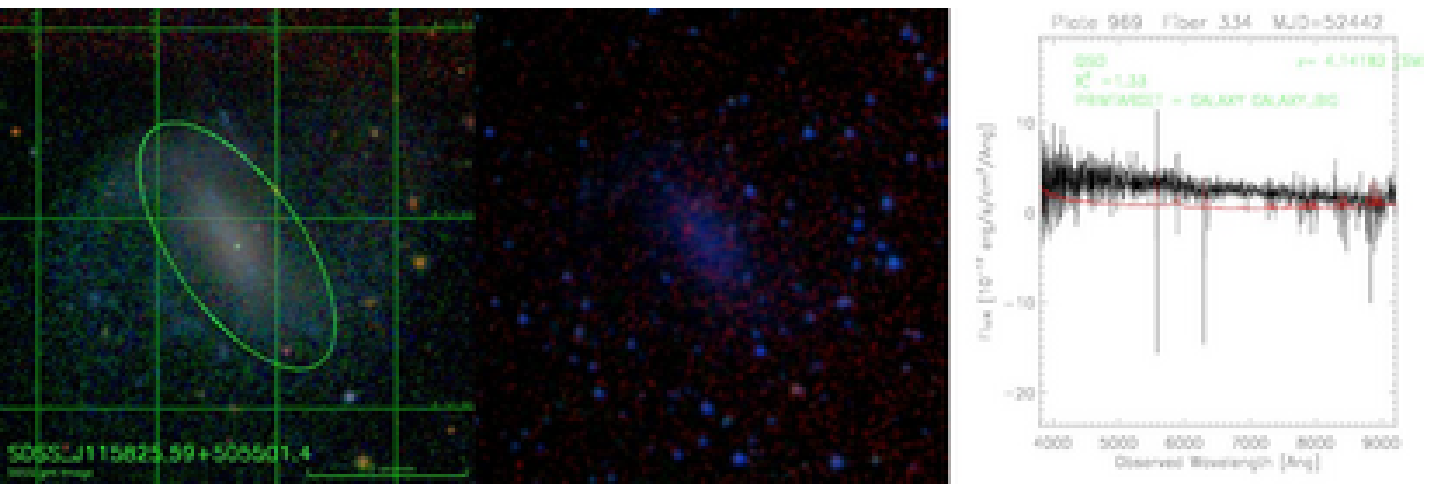,width=0.8\linewidth} \\
\epsfig{file=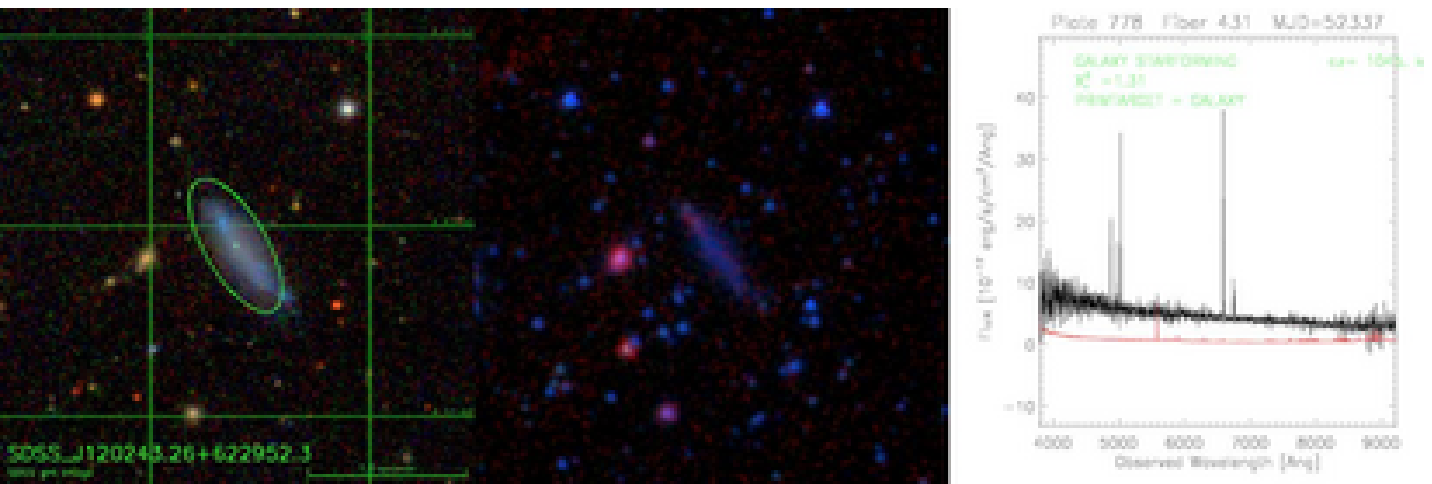,width=0.8\linewidth} \\
\epsfig{file=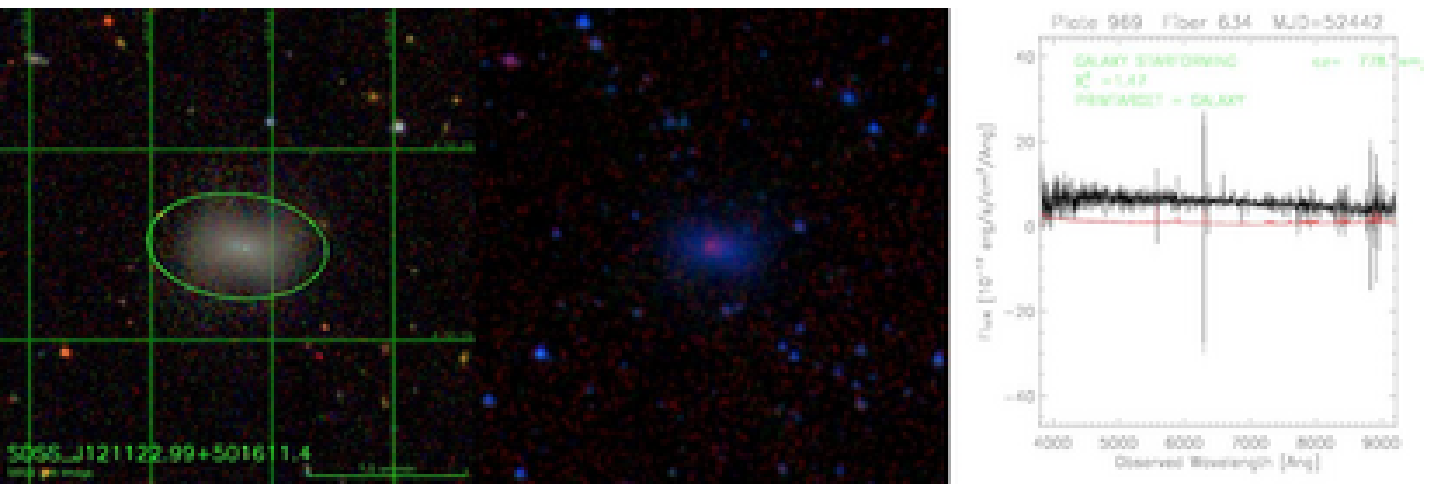,width=0.8\linewidth} \\
\epsfig{file=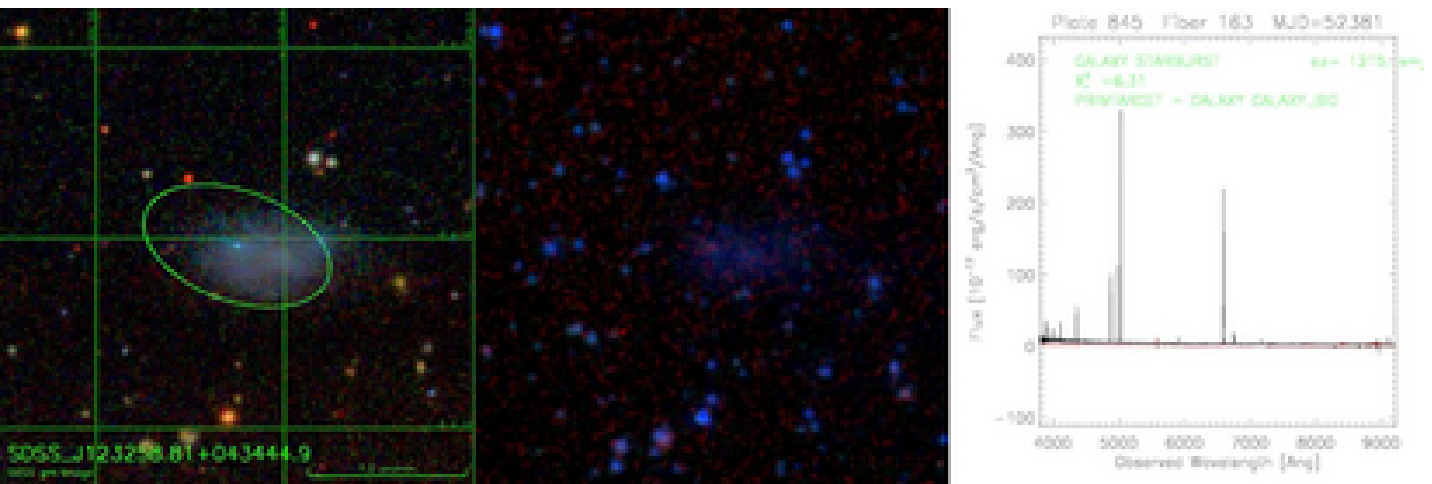,width=0.8\linewidth} \\
\end{tabular}
\figcaption{Cont.\label{galpic_2}}
\end{figure}
\addtocounter{figure}{-1}
\thispagestyle{empty}
\begin{figure}
\centering
\begin{tabular}{c}
\epsfig{file=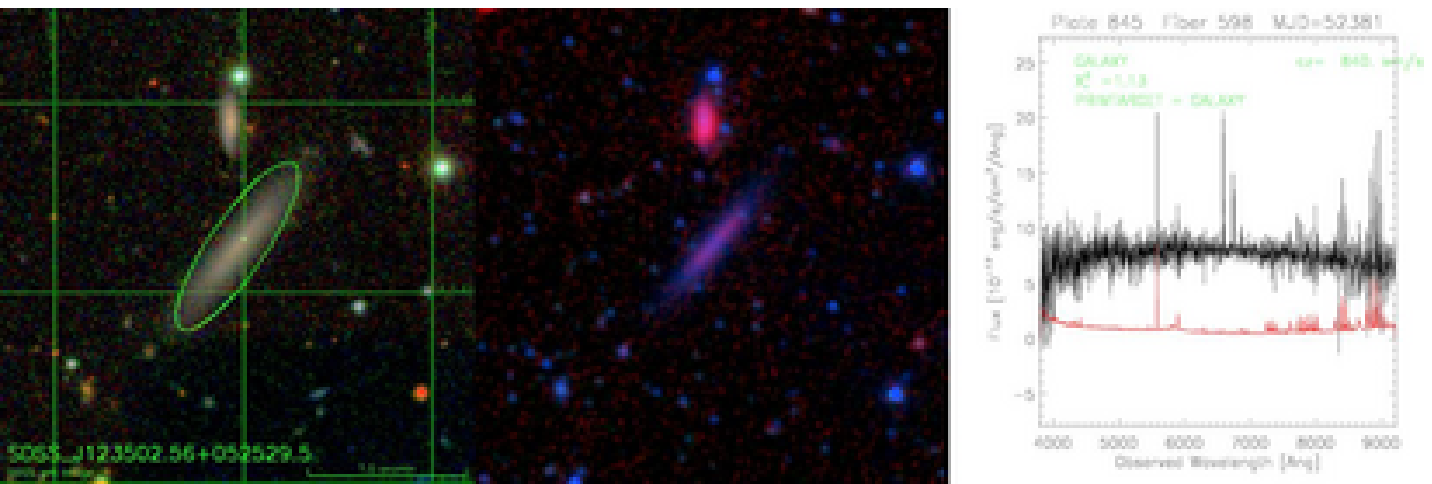,width=0.8\linewidth} \\
\epsfig{file=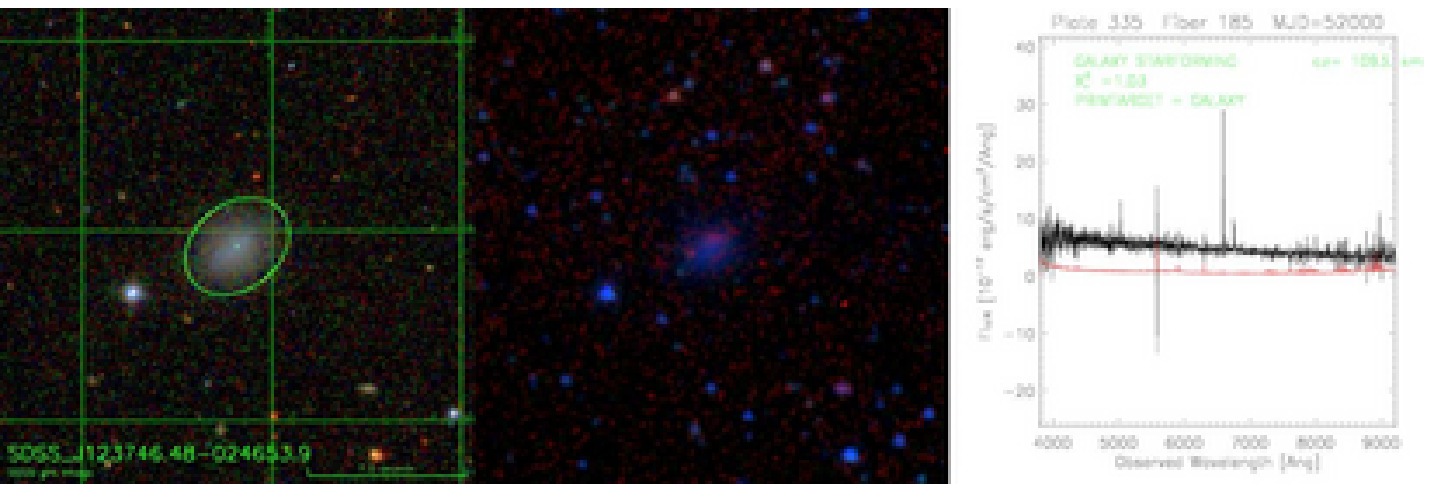,width=0.8\linewidth} \\
\epsfig{file=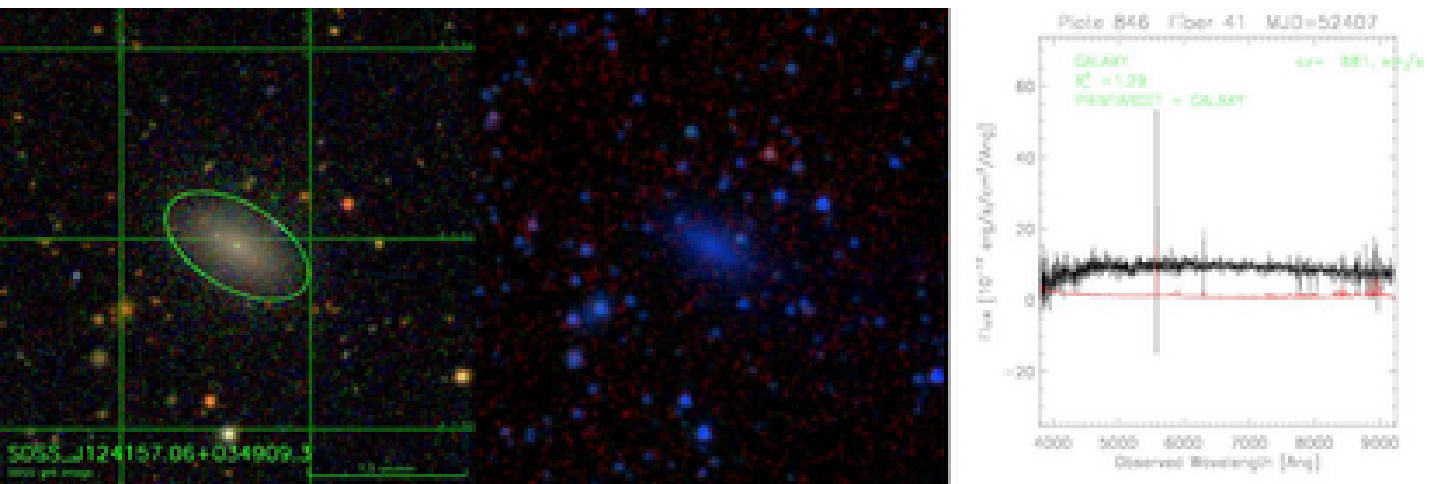,width=0.8\linewidth} \\
\epsfig{file=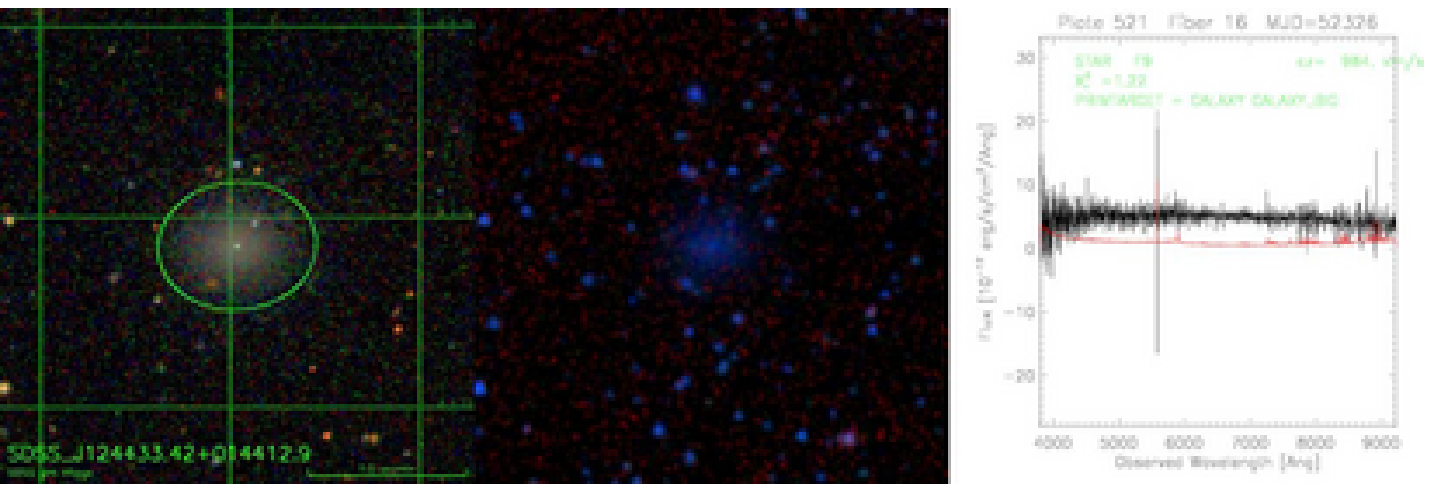,width=0.8\linewidth} \\
\epsfig{file=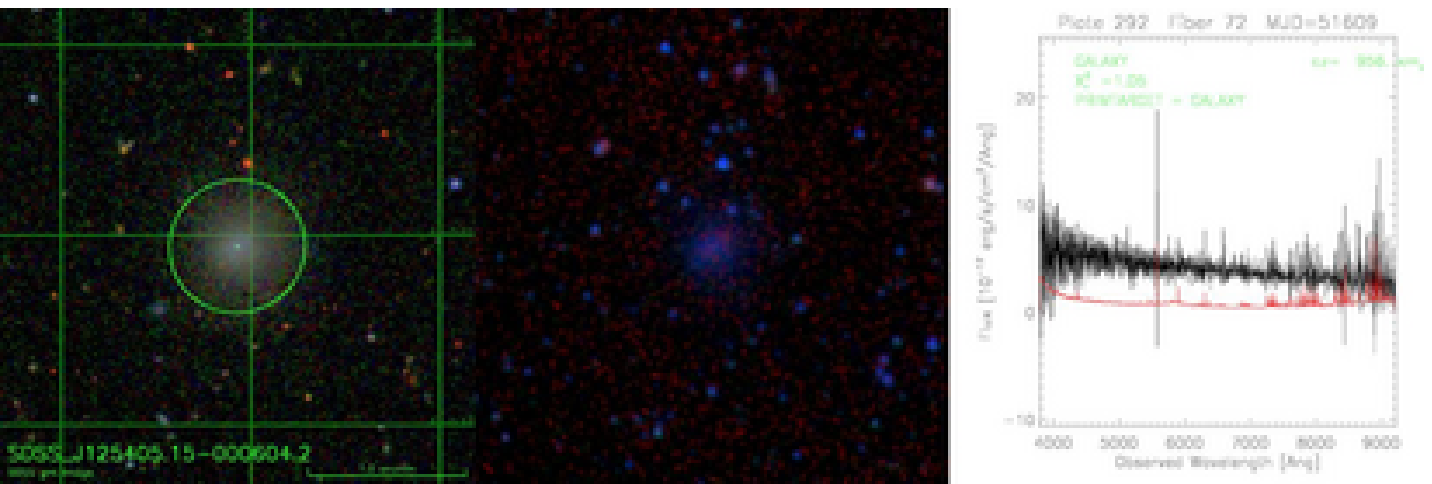,width=0.8\linewidth} \\
\end{tabular}
\figcaption{Cont.\label{galpic_3}}
\end{figure}
\addtocounter{figure}{-1}
\thispagestyle{empty}
\begin{figure}
\centering
\begin{tabular}{c}
\epsfig{file=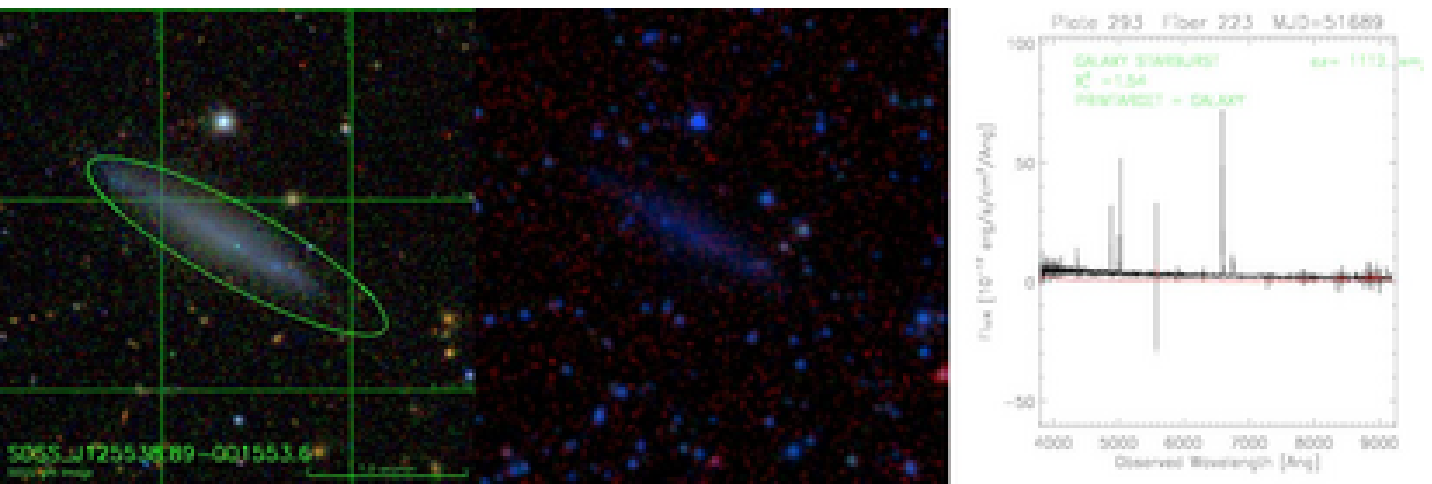,width=0.8\linewidth} \\
\epsfig{file=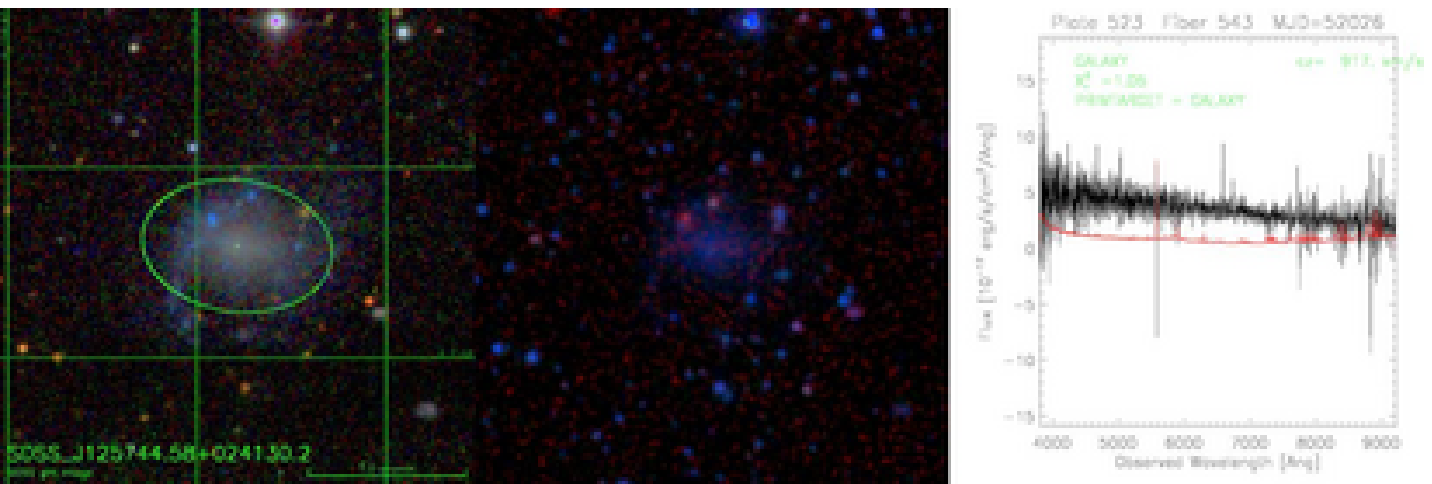,width=0.8\linewidth} \\
\epsfig{file=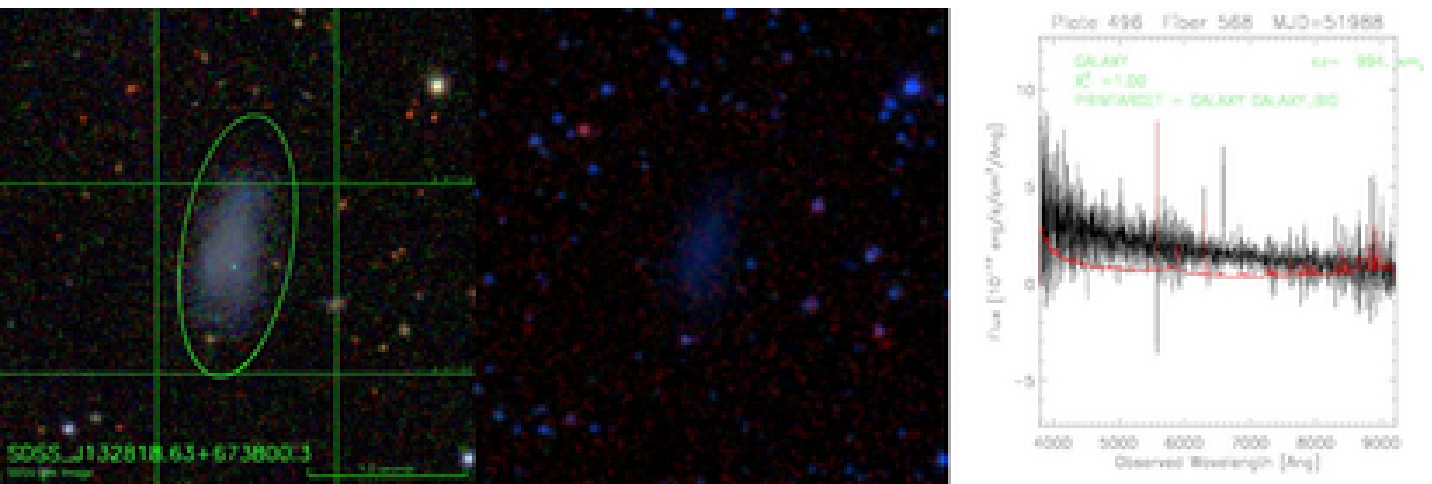,width=0.8\linewidth} \\
\epsfig{file=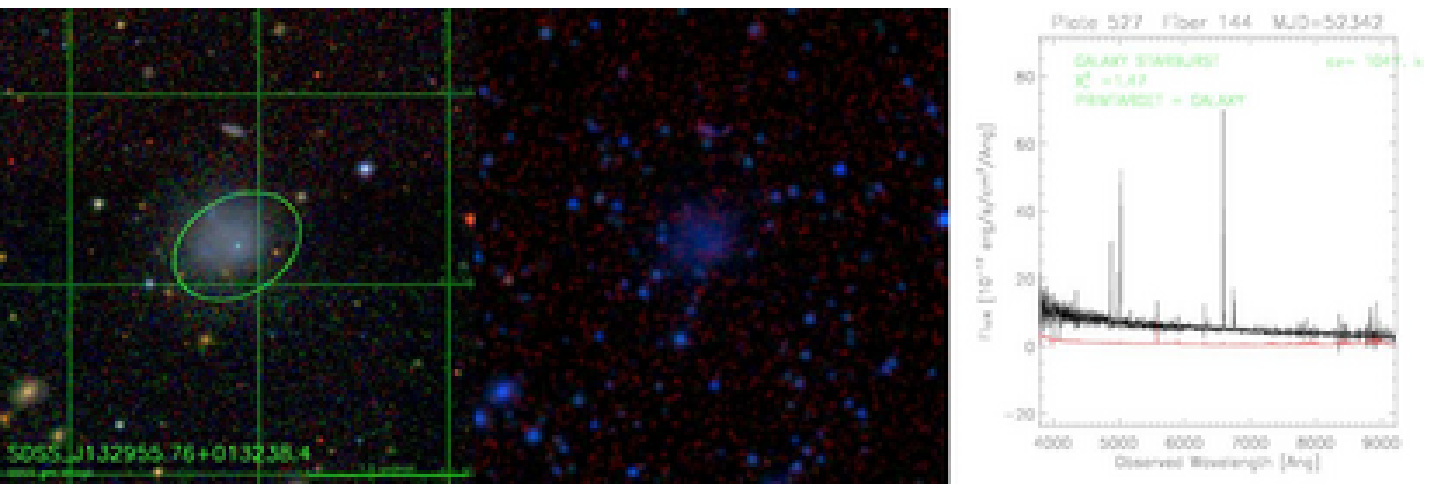,width=0.8\linewidth} \\
\epsfig{file=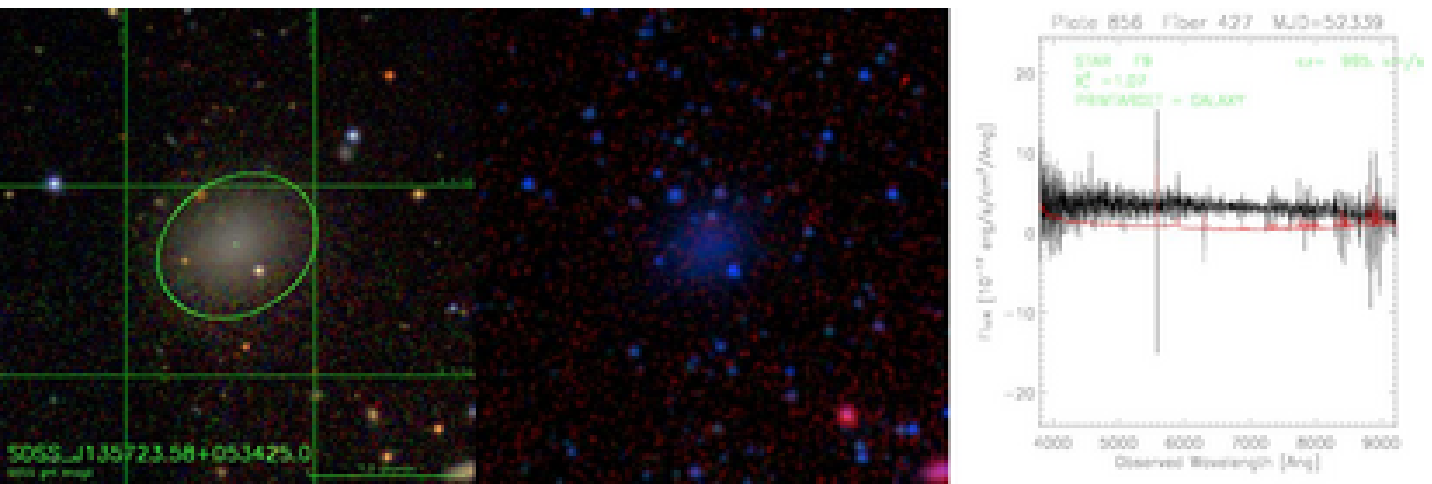,width=0.8\linewidth} \\
\end{tabular}
\figcaption{Cont.\label{galpic_4}}
\end{figure}
\addtocounter{figure}{-1}
\thispagestyle{empty}
\begin{figure}
\centering
\begin{tabular}{c}
\epsfig{file=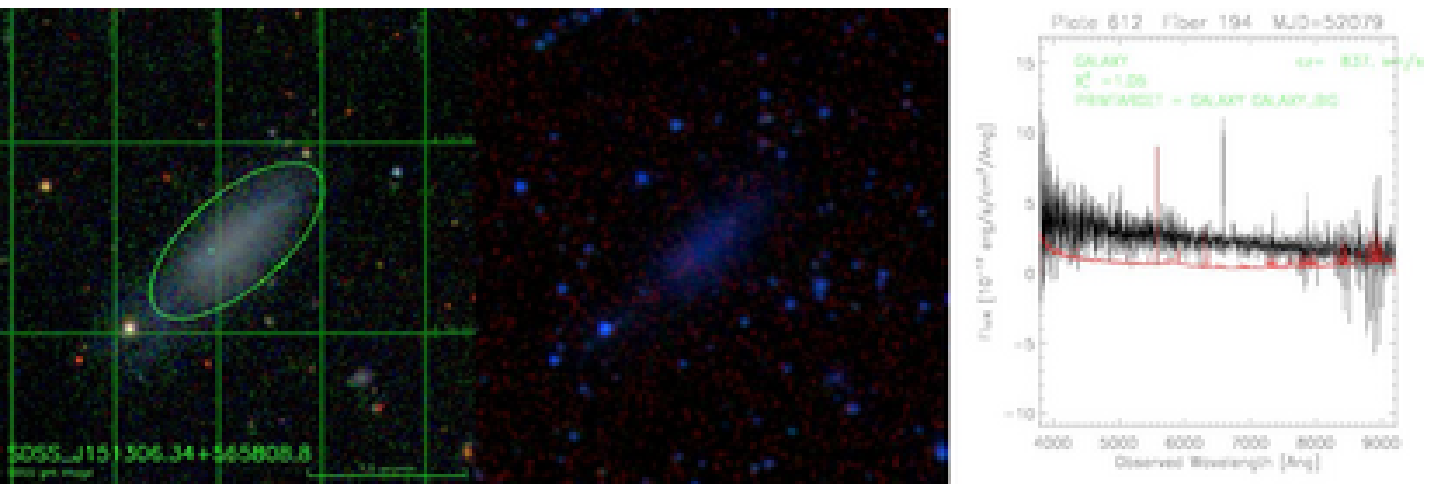,width=0.8\linewidth} \\
\epsfig{file=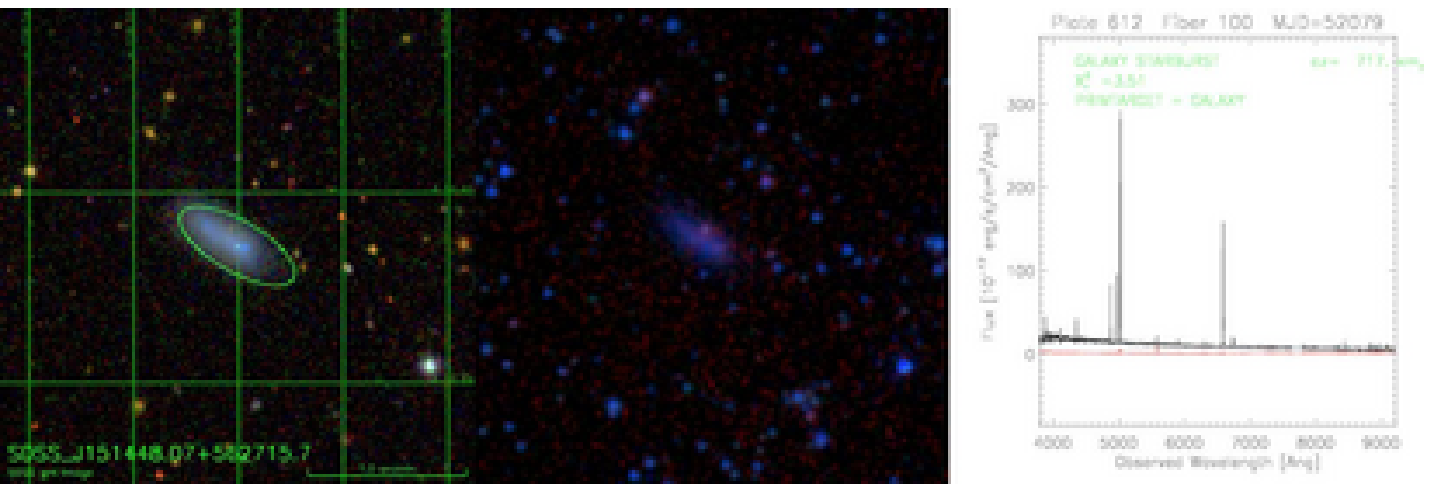,width=0.8\linewidth} \\
\epsfig{file=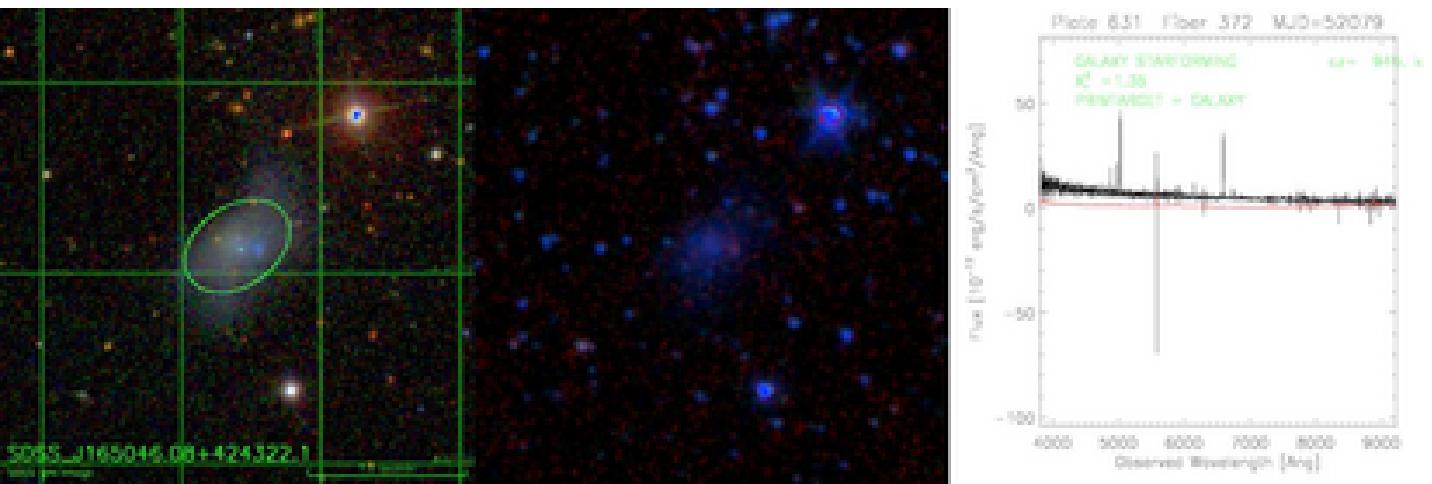,width=0.8\linewidth} \\
\epsfig{file=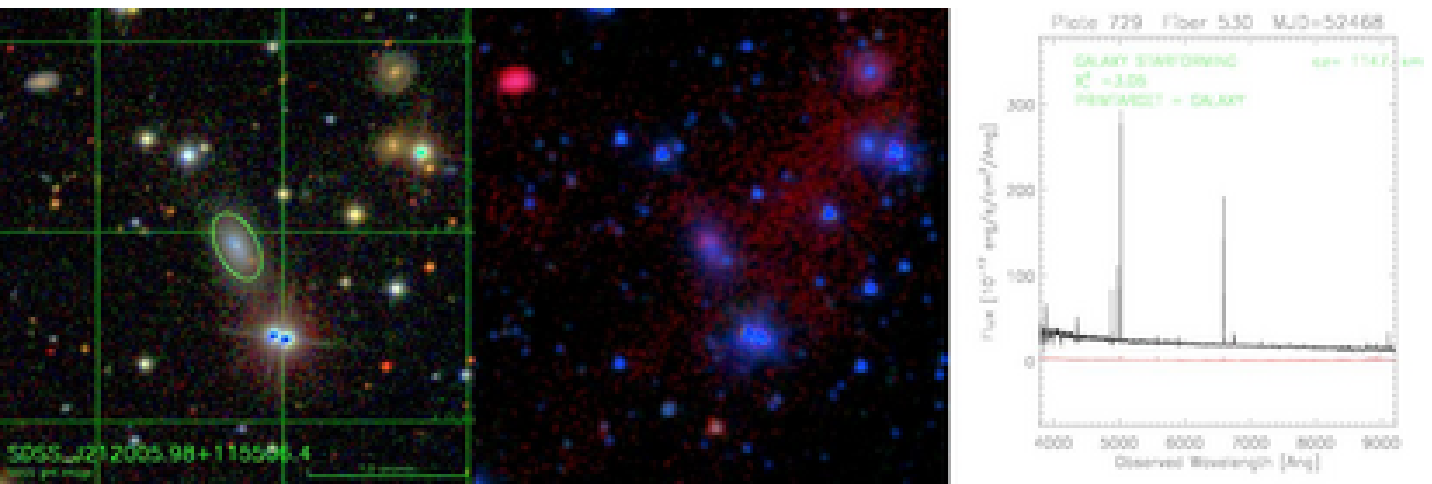,width=0.8\linewidth} \\
\epsfig{file=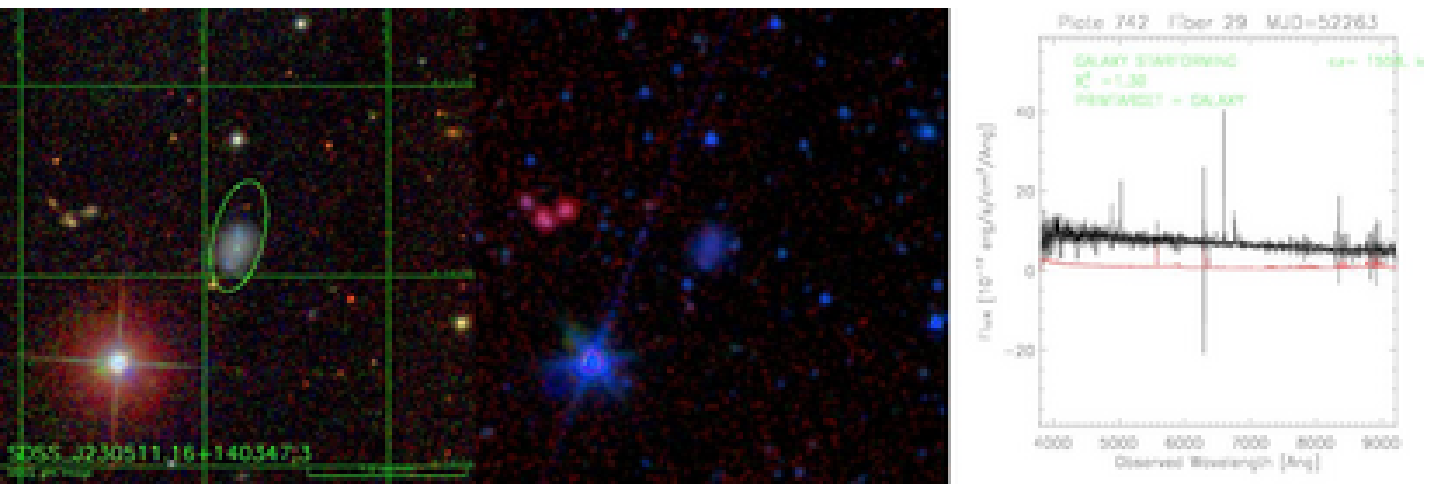,width=0.8\linewidth} \\
\end{tabular}
\figcaption{Cont.\label{galpic_5}}
\end{figure}

\begin{figure*}[b]
\plotone{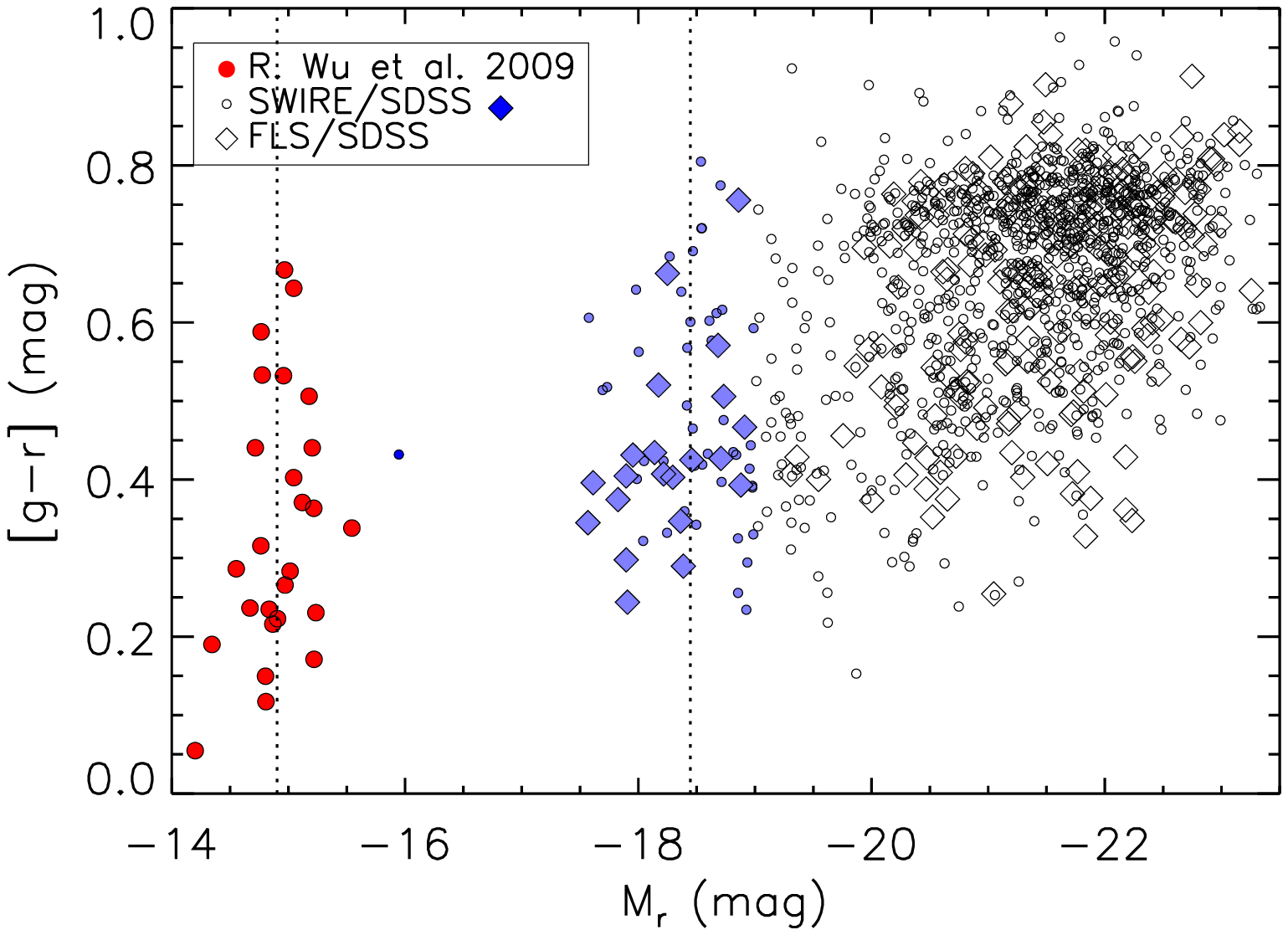}
\figcaption{The distribution of all galaxies studied in this work on the $[g-r]$ color and absolute magnitude $M_{r}$ plane. The symbol shapes indicate survey origin and the colors are used to indicate the $M_{r}$ range~(light blue: $-19.0<M_{r}<-17.5$, dark blue: $M_{r}>-17.5$). The 29 very low-luminosity galaxies studied in this work are indicated by the red circles. The two dotted lines indicate the median $M_{r}$ value of dwarf galaxies~($M_{r}>-19$) from the FLS/SDSS and SWIRE/SDSS galaxies~(left) and the median $M_{r}$ value of the very faint galaxies in our sample. The very low-luminosity galaxies presented in this work stand out clearly in this plot to be $\sim3$~\mag~fainter than the dwarfs included in the FLS/SDSS and SWIRE/SDSS samples.\label{color_absmr}}
\end{figure*}

\begin{figure*}
\plotone{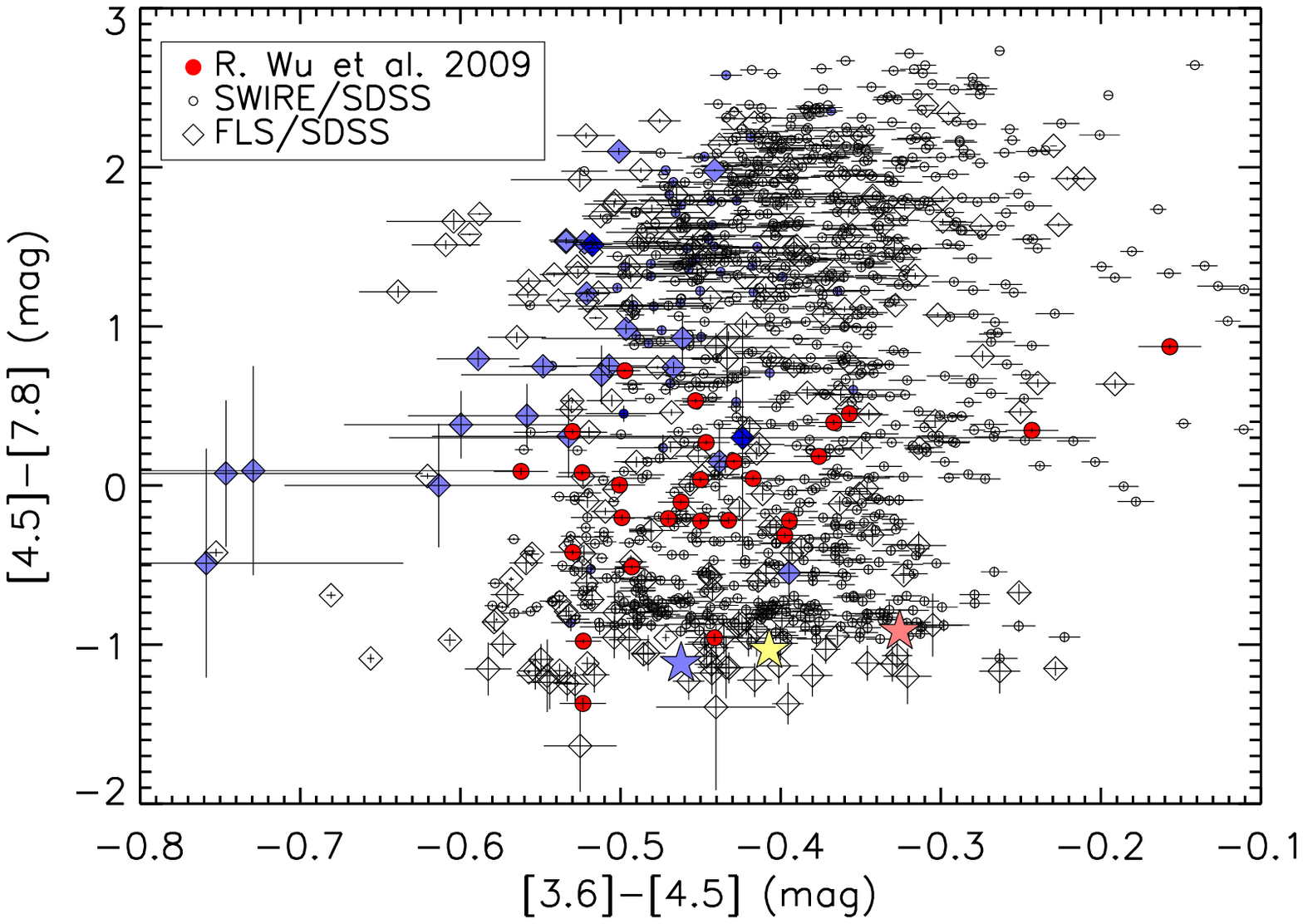}
\figcaption{Mid-infrared color--color diagram. Galaxies plotted are of $z<0.05$. The three over-plotted stars correspond to the predicted colors of three blackbody spectra of temperatures, $T_{1}=3000k$, $T_{2}=5000K$, $T_{3}=9000K$. Symbol colors as in Fig.(\ref{color_absmr}).\label{color_12_42}}
\end{figure*}

\begin{figure*}
\plotone{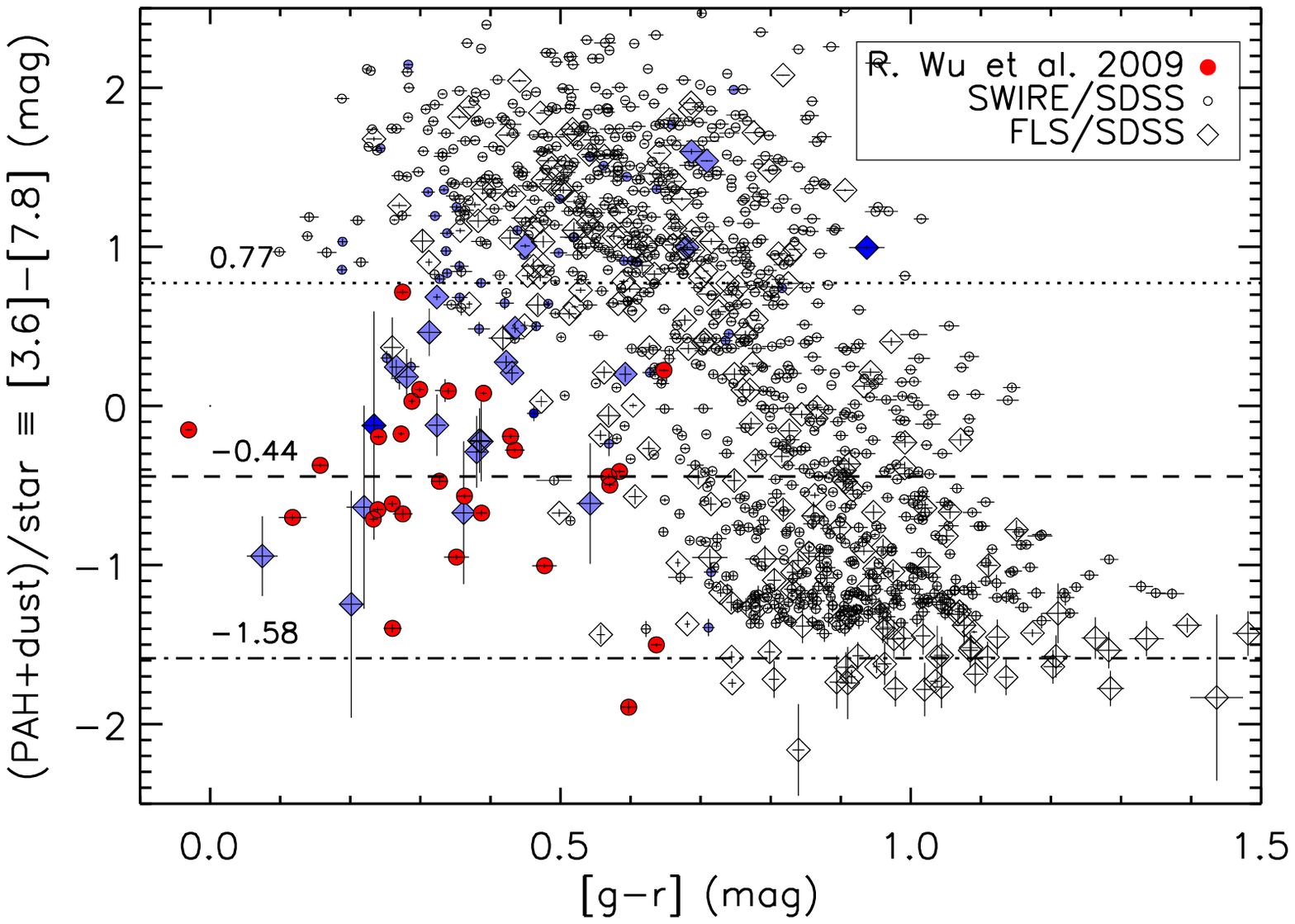}
\figcaption{MIR-optical color--color diagram. The dotted line indicates the median $\cha-\chd$ color for galaxies of intermediate luminosity, i.e. $-19<M_{r}<-17.5$\mag. The dashed line indicates the median $\cha-\chd$ color for galaxies of the lowest-luminosity, i.e. $M_{r}>-17.5$\mag. The dot-dashed line indicates the model-predicted $\cha-\chd$ color for a completely dust-free environment. Symbol colors as in Fig.(\ref{color_absmr}). Compared to more luminous galaxies of similar color, dwarf galaxies are clearly bluer in the $\cha-\chd$ color, which implies weaker PAH emission.\label{color_gr_14}}
\end{figure*}

\begin{figure*}
\plotone{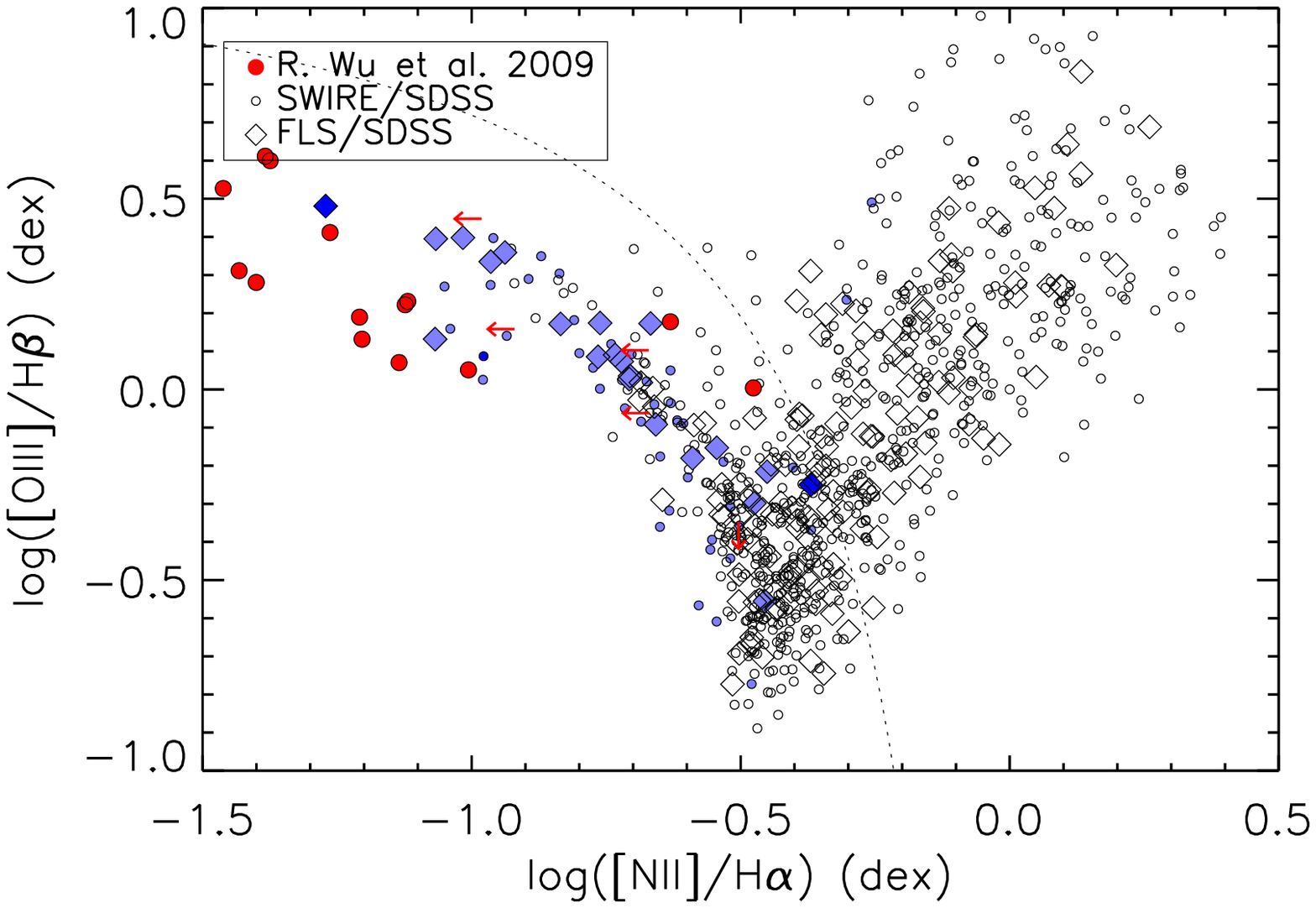}
\figcaption{The BPT diagram~\citep{baldwin-pasp-1981}. The dotted line is the demarcation curve separating star-forming galaxies (below) from AGNs (above)~\citep{kauffmann-mnras-2003}. Six very low-luminosity galaxies with S/N$<$3 are plotted as arrows indicating upper limits. Limits are calculated with  The leftward arrows indicates galaxies with low S/N in [N\textsc{ii}] and are shown with their measured [N\textsc{ii}] as an upper limit. The downward arrows indicate galaxies with low S/N in [O\textsc{ii}]  and are shown with their measured [O\textsc{iii}] as an upper limit. Symbol colors as in Fig.(\ref{color_absmr}).\label{o3_n2}}
\end{figure*}

\begin{figure*}
\epsscale{1.0}
\plotone{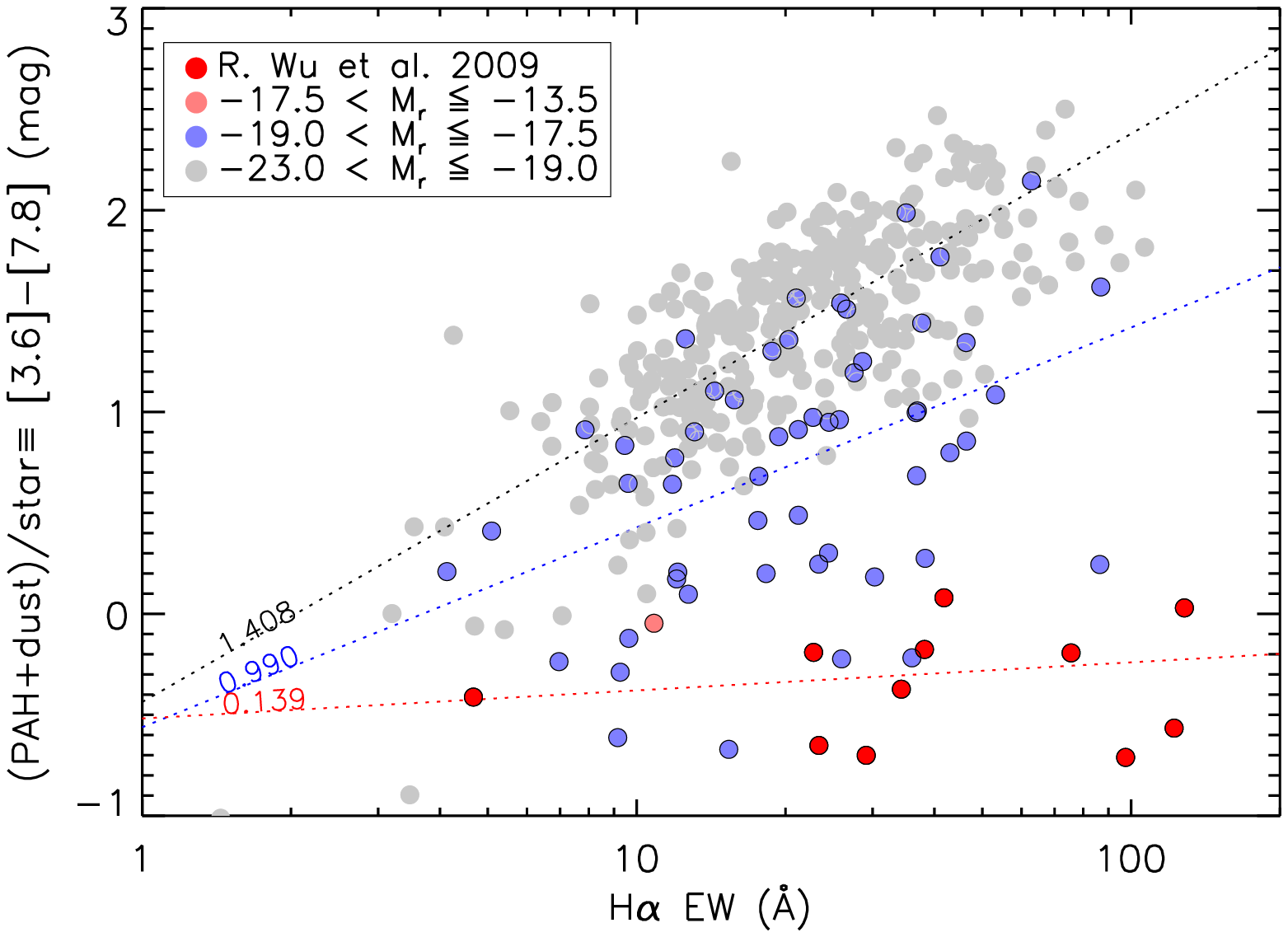}
\figcaption{Color--line-strength diagram for star-forming galaxies. The very faint galaxies presented in this work are highlighted with red colors, and they belong to the lowest-luminosity galaxy group in our analysis. For SWIRE/SDSS and FLS/SDSS galaxies: the gray color indicates galaxies of the highest luminosity, $-23.0<M_{r}\leq-19.0$\mag, the light blue color indicates galaxies of the intermediate luminosity, $-19.0<M_{r}\leq-17.5$\mag, and the pink color indicates galaxies of the lowest luminosity, $-17.5<M_{r}\leq-13.5$\mag. The black, blue and red dotted lines, with their slope labeled in the same color, indicate the linear regression for the highest, the intermediate, and the lowest-luminosity galaxies respectively. While the $\cha-\chd$ colors of luminous galaxies show a strong correlation with EW(H$\alpha$), the $\cha-\chd$ colors of the dwarf galaxies do not.\label{pah_haew}}
\end{figure*}

\begin{figure*}
\plotone{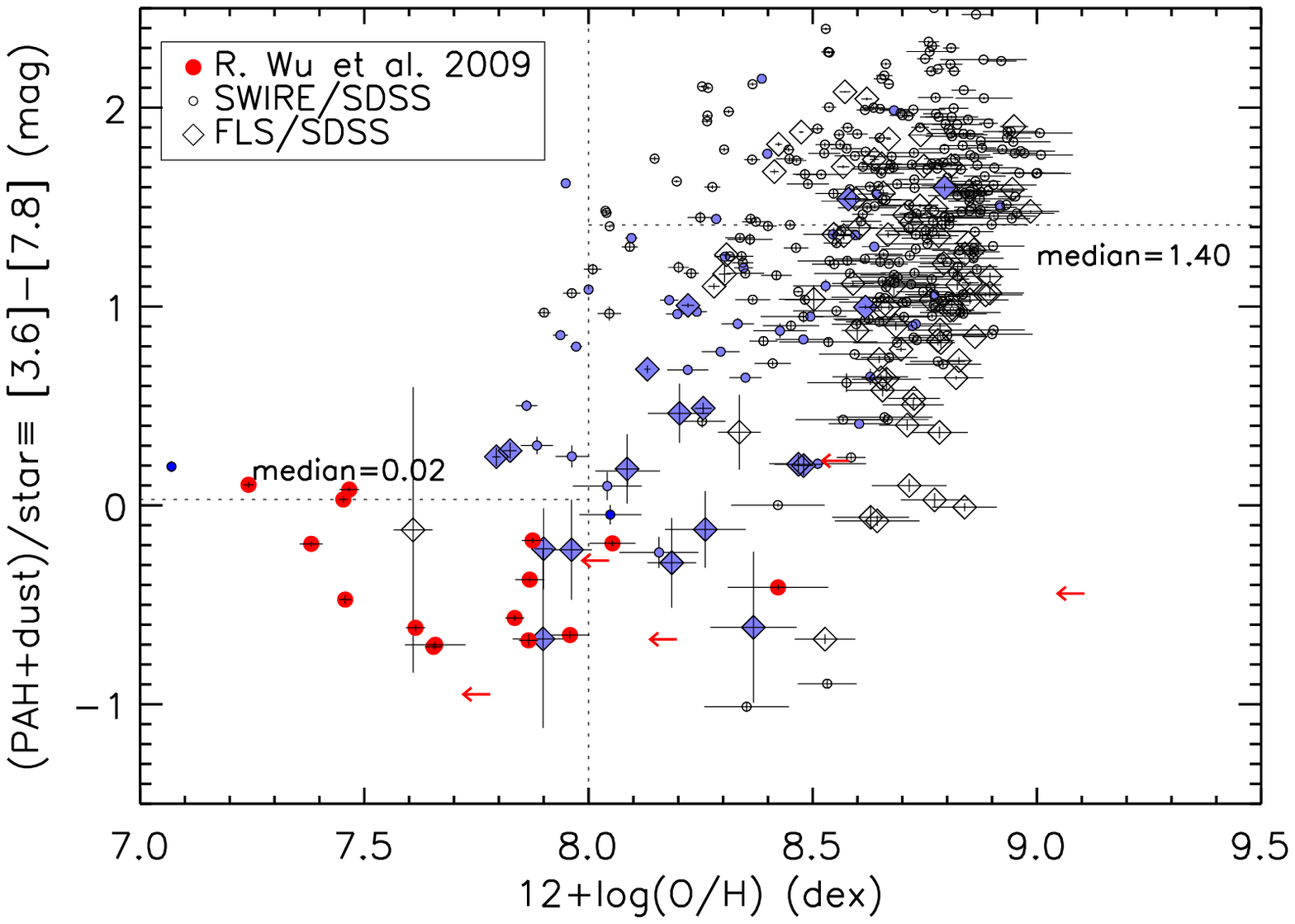}
\figcaption{MIR color--metallicity diagram for star-forming galaxies. The arrows indicate upper limits for the six very low-luminosity galaxies without good nebular line measurements. Symbol colors as in Fig.(\ref{color_absmr}). The lowest-luminosity galaxies~(red solid circles) show no clear dependence of $\cha-\chd$ on metallicity.\label{color_metal}}
\end{figure*}

\begin{figure*}
\plotone{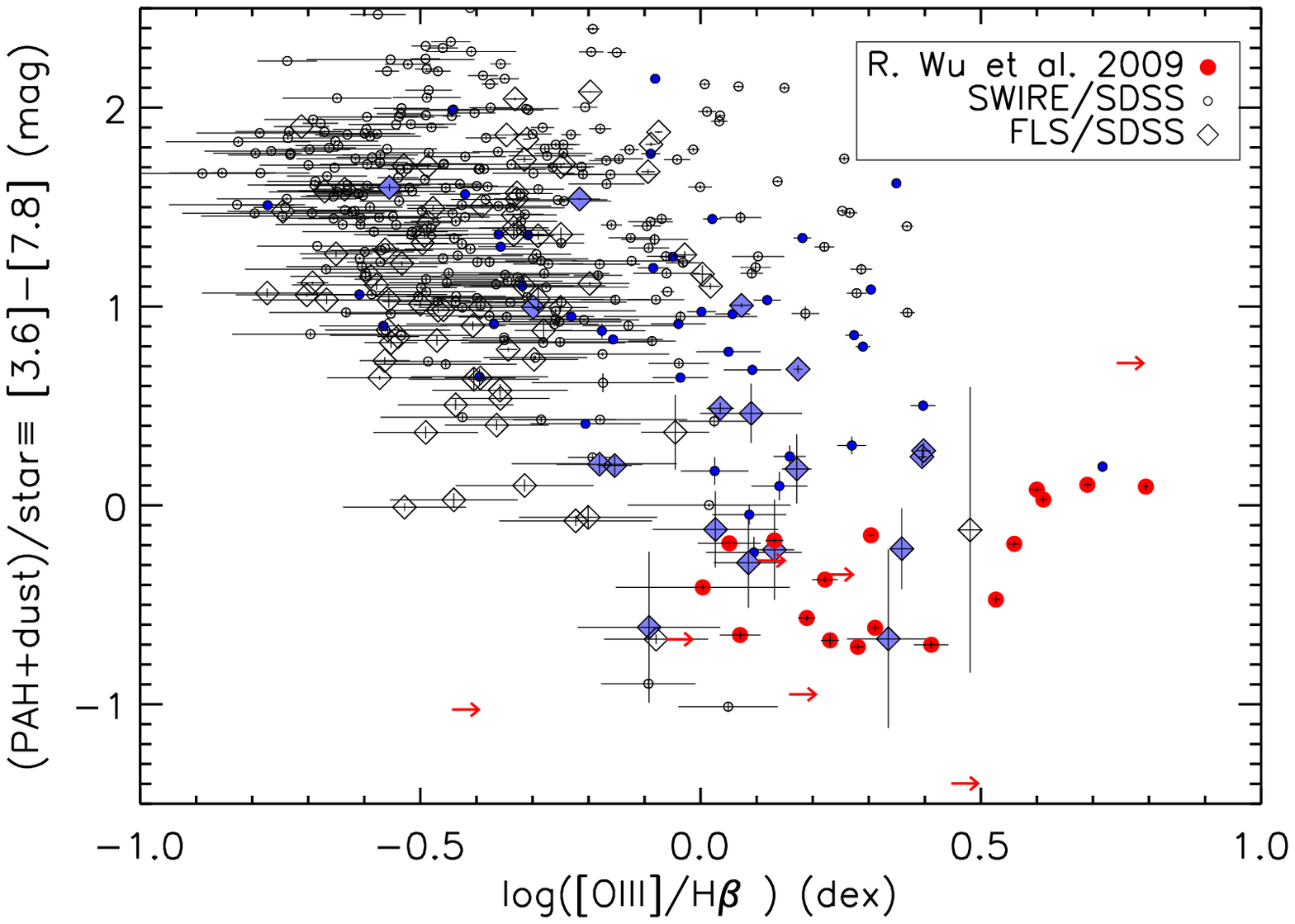}
\figcaption{MIR color--ionization parameter diagram for star-forming galaxies. Symbol colors as in Fig.(\ref{color_absmr}). The lowest-luminosity galaxies~(red solid circles) show no clear dependence of $\cha-\chd$ on the ionization parameter.\label{pah_o3}}
\end{figure*}

\begin{figure*}
\epsscale{1.0}
\plottwo{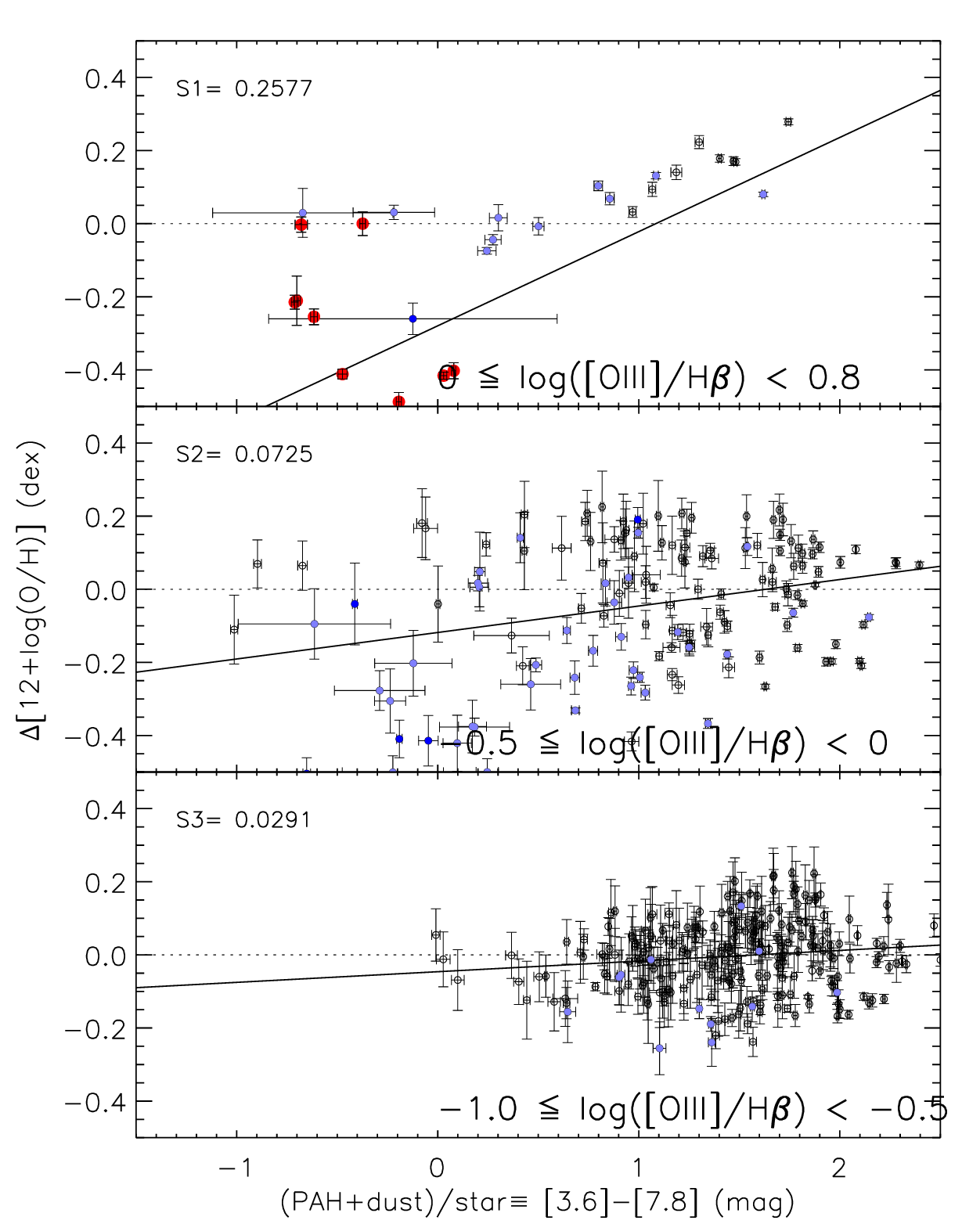}{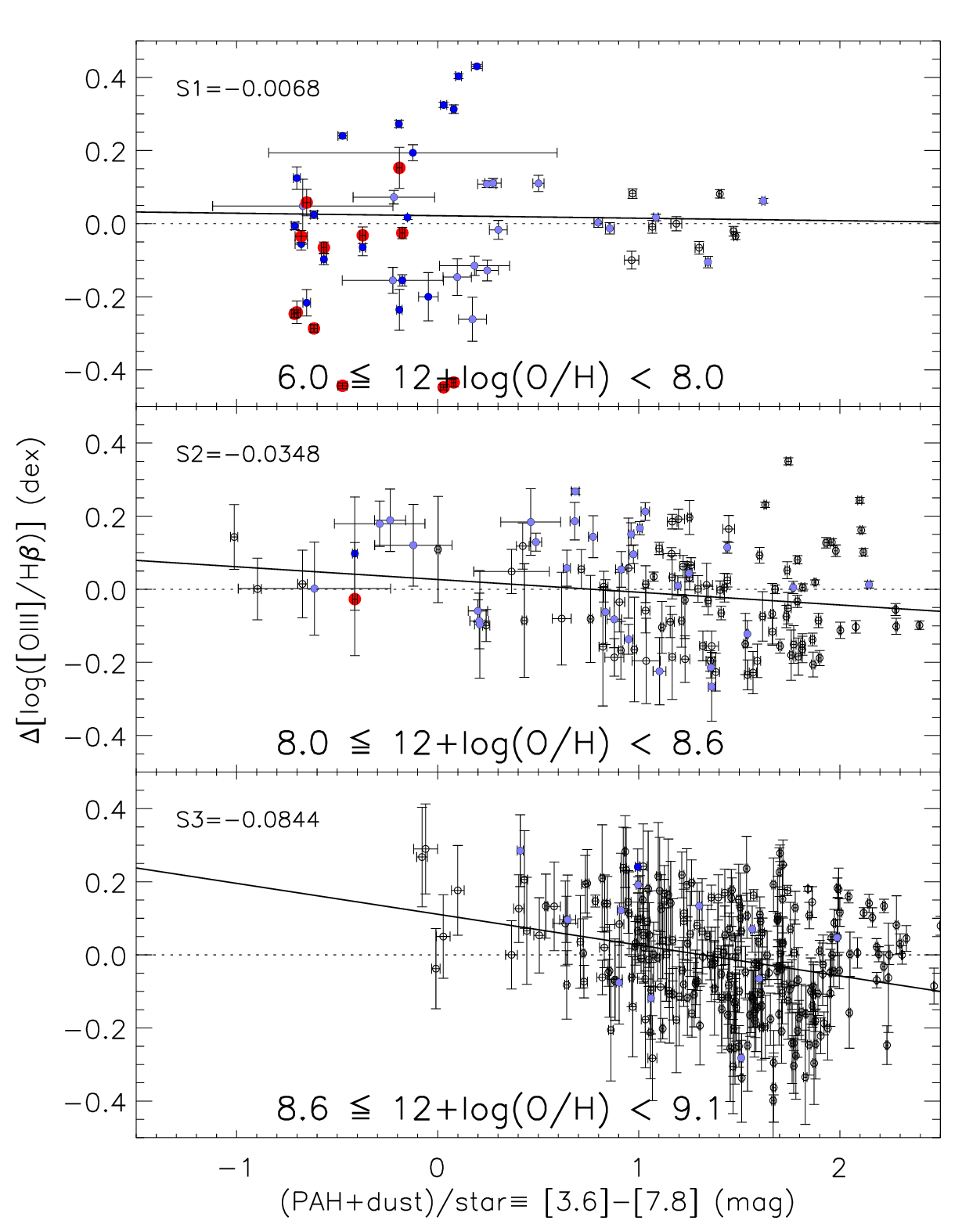}
\figcaption{The left plot shows relationship between $12+\log(O/H)$ and the $\cha-\chd$ color from harder~(top) to softer~(bottom) radiation fields. The right plot shows relationship between $\othree$~and PAH-dust/star from low metallicity~(top) to high metallicity~(bottom). In both plots, the galaxies are plotted according to their luminosity, the median luminosity is $M_{r}=-17.9$, $M_{r}=-20.0$, $M_{r}=-20.9$, from top to bottom respectively. The linear regression line is plotted in each panel as a solid line with its slope labeled at the top left corner. Symbol colors as in Fig.(\ref{color_absmr}). The $\cha-\chd$ colors of dwarf galaxies show a stronger dependence on metallicity than on radiation hardness.\label{irc_delta}}
\end{figure*}

\clearpage
\tiny
\renewcommand{\thefootnote}{\alph{footnote}}
\begin{landscape}
\begin{center}
\begin{longtable}{cccccccccc}
\caption[SDSS photometry and spectroscopic data]{SDSS photometry and spectroscopic data} \label{gals} \\
\hline
Name & redshift & $M_{r}$ & $g^{a}$ & $r$ & $i$ & $\ntwo$ & $\othree$ & $\log(H\alpha)$ & $\log(H\beta)$ \\
     &    & (\mag) & (\mujy) & (\mujy) & (\mujy) & (\dex) & (\dex) & (\dex) & (\dex)\\
\hline
SDSSJ024200.36$+$000052.3 & 0.0036 & $-14.96$ & $897.537\pm 1.495$ & $1407.72\pm 2.176$ & $1740.60\pm 3.242$ & $ 0.09\pm 0.138$ &$ 0.76\pm 0.333$ &$ 0.87\pm 0.333$ &$ 1.39\pm 0.102$  \\
SDSSJ024815.93$-$081716.5 & 0.0046 & $-14.90$ & $1127.41\pm 3.554$ & $1261.77\pm 5.424$ & $1087.39\pm 8.223$ & $ 0.33\pm 0.074$ &$ 2.53\pm 0.012$ &$ 4.63\pm 0.006$ &$ 6.93\pm 0.301$  \\
SDSSJ093402.00$+$551428.1 & 0.0024 & $-13.55$ & $891.043\pm 2.188$ & $806.152\pm 2.308$ & $514.100\pm 2.592$ & $ 0.10\pm 0.147$ &$ 1.38\pm 0.086$ &$ 3.79\pm 0.010$ &$ 3.91\pm 0.009$  \\
SDSSJ102701.74$+$561614.4 & 0.0028 & $-14.87$ & $2101.39\pm 3.703$ & $2513.97\pm 4.639$ & $2742.25\pm 5.855$ & $ 0.32\pm 0.019$ &$ 1.42\pm 0.084$ &$ 2.27\pm 0.025$ &$ 2.62\pm 0.013$  \\
SDSSJ105539.17$+$022344.7 & 0.0034 & $-14.81$ & $1459.97\pm 4.295$ & $1510.87\pm 5.729$ & $1307.30\pm 8.779$ & $ 0.44\pm 0.074$ &$ 1.88\pm 0.036$ &$ 3.70\pm 0.010$ &$ 3.51\pm 0.010$  \\
SDSSJ111054.19$+$010530.4 & 0.0033 & $-14.80$ & $1589.49\pm 4.348$ & $1831.68\pm 6.193$ & $1844.86\pm 10.93$ & $ 0.44\pm 0.065$ &$ 1.43\pm 0.080$ &$ 2.74\pm 0.014$ &$ 2.86\pm 0.011$  \\
SDSSJ115132.93$-$022222.0 & 0.0034 & $-14.20$ & $1265.14\pm 3.270$ & $1170.21\pm 4.307$ & $910.937\pm 6.351$ & $ 0.26\pm 0.052$ &\nodata$\pm$ \nodata$^{b}$ &$ 4.68\pm 0.008$ &\nodata$\pm$ \nodata$^{b}$  \\
SDSSJ115314.10$-$032432.2 & 0.0041 & $-15.24$ & $1208.62\pm 3.006$ & $1364.37\pm 3.606$ & $1418.82\pm 5.413$ & $ 0.16\pm 0.090$ &$ 2.06\pm 0.033$ &$ 3.52\pm 0.009$ &$ 3.45\pm 0.009$  \\
SDSSJ115735.29$+$021004.1 & 0.0031 & $-14.76$ & $1200.20\pm 4.142$ & $1584.92\pm 5.752$ & $1889.93\pm 9.023$ & $ 0.40\pm 0.058$ &$ 0.74\pm 0.370$ &$ 1.14\pm 0.230$ &$ 1.64\pm 0.065$  \\
SDSSJ115747.56$+$531404.2 & 0.0037 & $-15.22$ & $1522.41\pm 4.155$ & $1771.43\pm 5.609$ & $1865.45\pm 9.531$ & $ 0.47\pm 0.048$ &$ 1.45\pm 0.068$ &$ 2.65\pm 0.015$ &$ 2.85\pm 0.011$  \\
SDSSJ115825.59$+$505501.4& \nodata$^{c}$  & \nodata$^{c}$ & $2056.72\pm 6.677$ & $3031.81\pm 9.896$ & $3305.68\pm 15.68$ & \nodata$\pm$\nodata$^{c}$ &\nodata$\pm$ \nodata$^{c}$ &\nodata$\pm$ \nodata$^{c}$ &\nodata$\pm$ \nodata$^{c}$  \\
SDSSJ120243.26$+$622952.3 & 0.0035 & $-14.84$ & $1275.99\pm 3.967$ & $1606.47\pm 6.087$ & $1797.42\pm 9.485$ & $ 0.41\pm 0.037$ &$ 1.07\pm 0.112$ &$ 1.93\pm 0.032$ &$ 2.19\pm 0.016$  \\
SDSSJ121122.99$+$501611.4 & 0.0026 & $-14.78$ & $1618.25\pm 5.161$ & $2612.99\pm 7.605$ & $3439.02\pm 12.91$ & $ 0.57\pm 0.034$ &$ 1.05\pm 0.191$ &$ 0.99\pm 0.282$ &$ 1.53\pm 0.069$  \\
SDSSJ123258.81$+$043444.9 & 0.0041 & $-14.97$ & $1200.80\pm 5.651$ & $1437.49\pm 8.151$ & $1426.35\pm 12.94$ & $ 0.57\pm 0.028$ &$ 1.54\pm 0.047$ &$ 3.04\pm 0.011$ &$ 3.01\pm 0.009$  \\
SDSSJ123502.56$+$052529.5 & 0.0028 & $-14.97$ & $1425.11\pm 4.707$ & $2626.60\pm 6.971$ & $3591.99\pm 11.61$ & $ 0.41\pm 0.053$ &$ 1.30\pm 0.106$ &$ 0.81\pm 0.362$ &$ 1.88\pm 0.039$  \\
SDSSJ123746.48$-$024653.9 & 0.0036 & $-15.05$ & $1158.10\pm 4.155$ & $1644.46\pm 6.411$ & $1929.47\pm 10.37$ & $ 0.37\pm 0.109$ &$ 1.07\pm 0.141$ &$ 1.46\pm 0.093$ &$ 2.08\pm 0.025$  \\
SDSSJ124157.06$+$034909.3 & 0.0029 & $-15.05$ & $1374.51\pm 5.865$ & $2468.44\pm 7.623$ & $3358.30\pm 11.27$ & $ 0.41\pm 0.074$ &\nodata$\pm$ \nodata$^{d}$ &\nodata$\pm$ \nodata$^{d}$ &$ 0.59\pm 0.654$  \\
SDSSJ124433.42$+$014412.9 & 0.0033 & $-14.77$ & $1007.38\pm 5.955$ & $1737.18\pm 9.712$ & $2249.55\pm 20.62$ & $ 0.70\pm 0.026$ &$-0.03\pm 3.993$ &$ 0.88\pm 0.766$ &\nodata$\pm$ \nodata$^{d}$  \\
SDSSJ125405.15$-$000604.2 & 0.0032 & $-14.72$ & $1095.63\pm 4.844$ & $1745.71\pm 7.460$ & $2066.04\pm 11.09$ & $ 0.42\pm 0.053$ &$ 0.92\pm 0.218$ &$-0.01\pm 1.971$ &$ 1.35\pm 0.092$  \\
SDSSJ125538.89$-$001553.6 & 0.0037 & $-15.01$ & $1660.26\pm 5.904$ & $2184.93\pm 8.746$ & $2335.07\pm 13.14$ & $ 0.51\pm 0.044$ &$ 1.31\pm 0.072$ &$ 2.19\pm 0.022$ &$ 2.52\pm 0.012$  \\
SDSSJ125744.58$+$024130.2 & 0.0031 & $-15.12$ & $2240.52\pm 6.990$ & $2948.77\pm 10.59$ & $3479.96\pm 18.28$ & $ 0.68\pm 0.033$ &$ 0.55\pm 0.529$ &$ 1.26\pm 0.150$ &$ 1.52\pm 0.072$  \\
SDSSJ132818.63$+$673800.3 & 0.0033 & $-14.67$ & $1381.99\pm 6.850$ & $1718.35\pm 10.42$ & $2018.72\pm 15.29$ & $ 0.70\pm 0.015$ &\nodata$\pm$ \nodata$^{d}$ &$ 1.04\pm 0.212$ &$ 1.36\pm 0.076$  \\
SDSSJ132955.76$+$013238.4 & 0.0035 & $-14.91$ & $1364.36\pm 4.113$ & $1729.62\pm 5.690$ & $1936.06\pm 11.04$ & $ 0.43\pm 0.058$ &$ 1.37\pm 0.070$ &$ 2.22\pm 0.023$ &$ 2.49\pm 0.013$  \\
SDSSJ135723.58$+$053425.0 & 0.0033 & $-15.18$ & $1565.85\pm 4.723$ & $2698.57\pm 6.906$ & $3679.03\pm 10.32$ & $ 0.46\pm 0.052$ &\nodata$\pm$ \nodata$^{d}$ &$ 0.84\pm 0.515$ &$ 0.74\pm 0.347$  \\
SDSSJ151306.34$+$565808.8 & 0.0028 & $-15.22$ & $1502.36\pm 5.795$ & $2142.83\pm 9.087$ & $2320.23\pm 12.32$ & $ 0.65\pm 0.017$ &$ 0.26\pm 0.837$ &$ 1.07\pm 0.185$ &$ 1.58\pm 0.052$  \\
SDSSJ151448.07$+$562715.7 & 0.0024 & $-14.34$ & $1844.21\pm 4.057$ & $2059.32\pm 5.649$ & $2074.61\pm 9.675$ & $ 0.45\pm 0.020$ &$ 1.29\pm 0.095$ &$ 2.94\pm 0.014$ &$ 2.84\pm 0.012$  \\
SDSSJ165046.08$+$424322.1 & 0.0032 & $-14.55$ & $1156.12\pm 3.620$ & $1531.40\pm 5.592$ & $1759.99\pm 8.323$ & $ 0.40\pm 0.043$ &$ 0.91\pm 0.251$ &$ 2.14\pm 0.034$ &$ 2.18\pm 0.024$  \\
SDSSJ212005.98$+$115506.4 & 0.0038 & $-15.54$ & $1869.63\pm 3.714$ & $2340.51\pm 5.045$ & $2601.28\pm 7.201$ & $ 0.36\pm 0.209$ &$ 1.61\pm 0.080$ &$ 3.04\pm 0.014$ &$ 2.98\pm 0.012$  \\
SDSSJ230511.16$+$140347.3 & 0.0052 & $-15.20$ & $1536.28\pm 3.826$ & $1798.14\pm 5.458$ & $1855.88\pm 8.372$ & $ 0.38\pm 0.459$ &$ 1.27\pm 0.141$ &$ 1.97\pm 0.060$ &$ 2.41\pm 0.020$  \\
\hline
\end{longtable}
\end{center}
\end{landscape}
\addtocounter{table}{-1}
\begin{center}
\begin{longtable}{ccccc}
\caption[\it{Cont.} IRAC photometry data]{\it{Cont.} IRAC photometry data}\\
\hline
Name & $\cha^{a}$ & $\chb$ & $\chc$ & $\chd$ \\
     &  (\mujy) & (\mujy) & (\mujy) & (\mujy) \\
\hline
SDSSJ024200.36$+$000052.3 & $911.803\pm 9.031$ &$574.986\pm 6.289$ &$588.335\pm 8.610$ &$576.740\pm 5.785$  \\
SDSSJ024815.93$-$081716.5 & $658.157\pm 8.740$ &$473.701\pm 6.499$ &$387.942\pm 6.366$ &$717.833\pm 8.061$  \\
SDSSJ093402.00$+$551428.1 & $135.009\pm 0.000$ &$138.013\pm 0.000$ &$116.173\pm 0.000$ &$117.531\pm 0.000$  \\
SDSSJ102701.74$+$561614.4 & $1214.86\pm 11.56$ &$805.237\pm 7.843$ &$757.430\pm 12.30$ &$1032.82\pm 6.367$  \\
SDSSJ105539.17$+$022344.7 & $687.935\pm 7.699$ &$549.941\pm 5.627$ &$439.309\pm 5.395$ &$756.773\pm 5.228$  \\
SDSSJ111054.19$+$010530.4 & $834.482\pm 8.413$ &$580.215\pm 5.676$ &$462.334\pm 9.035$ &$473.372\pm 5.593$  \\
SDSSJ115132.93$-$022222.0 & $410.484\pm 5.473$ &$355.238\pm 4.325$ &$396.437\pm 3.838$ &$793.324\pm 4.877$  \\
SDSSJ115314.10$-$032432.2 & $625.712\pm 6.857$ &$446.319\pm 5.258$ &$447.375\pm 5.638$ &$642.908\pm 5.868$  \\
SDSSJ115735.29$+$021004.1 & $784.686\pm 8.226$ &$518.469\pm 5.754$ &$382.643\pm 5.799$ &$421.999\pm 4.752$  \\
SDSSJ115747.56$+$531404.2 & $839.594\pm 9.364$ &$582.207\pm 6.617$ &$401.965\pm 7.872$ &$436.128\pm 4.709$  \\
SDSSJ115825.59$+$505501.4 & $1949.09\pm 14.24$ &$1187.81\pm 8.955$ &$907.450\pm 18.85$ &$1462.50\pm 6.108$  \\
SDSSJ120243.26$+$622952.3 & $850.689\pm 9.397$ &$579.285\pm 6.416$ &$374.683\pm 4.004$ &$603.201\pm 3.647$  \\
SDSSJ121122.99$+$501611.4 & $1853.59\pm 14.32$ &$1224.61\pm 12.56$ &$807.164\pm 12.93$ &$1267.80\pm 10.24$  \\
SDSSJ123258.81$+$043444.9 & $804.197\pm 8.721$ &$479.207\pm 5.377$ &$521.206\pm 12.13$ &$519.888\pm 9.406$  \\
SDSSJ123502.56$+$052529.5 & $2779.26\pm 16.20$ &$1758.06\pm 11.60$ &$2059.11\pm 16.02$ &$3416.55\pm 12.49$  \\
SDSSJ123746.48$-$024653.9 & $934.820\pm 9.519$ &$573.740\pm 6.291$ &$553.667\pm 8.051$ &$784.120\pm 5.906$  \\
SDSSJ124157.06$+$034909.3 & $2032.69\pm 13.83$ &$1255.39\pm 9.454$ &$1215.29\pm 12.22$ &$509.886\pm 3.588$  \\
SDSSJ124433.42$+$014412.9 & $1196.35\pm 11.06$ &$738.717\pm 7.031$ &$490.699\pm 12.25$ &$209.123\pm 7.066$  \\
SDSSJ125405.15$-$000604.2 & $1126.13\pm 10.72$ &$695.028\pm 6.697$ &$620.150\pm 11.48$ &$748.695\pm 6.833$  \\
SDSSJ125538.89$-$001553.6 & $1127.74\pm 10.62$ &$736.750\pm 6.759$ &$428.724\pm 10.00$ &$669.436\pm 4.939$  \\
SDSSJ125744.58$+$024130.2 & $1675.87\pm 12.66$ &$1128.75\pm 8.349$ &$1052.99\pm 12.40$ &$1297.97\pm 8.566$  \\
SDSSJ132818.63$+$673800.3 & $899.354\pm 10.44$ &$598.995\pm 7.083$ &\nodata$\pm$ \nodata$^{e}$&$248.100\pm 5.110$  \\
SDSSJ132955.76$+$013238.4 & $922.935\pm 9.173$ &$598.582\pm 5.961$ &$600.912\pm 10.82$ &$493.975\pm 12.95$  \\
SDSSJ135723.58$+$053425.0 & $1859.89\pm 12.70$ &$1181.21\pm 8.187$ &$721.603\pm 6.221$ &$737.474\pm 6.729$  \\
SDSSJ151306.34$+$565808.8 & $1246.93\pm 11.30$ &$765.442\pm 6.967$ &$608.073\pm 10.68$ &$519.772\pm 4.651$  \\
SDSSJ151448.07$+$562715.7 & $932.387\pm 9.294$ &$659.415\pm 6.810$ &$588.227\pm 7.865$ &$780.306\pm 4.317$  \\
SDSSJ165046.08$+$424322.1 & $923.903\pm 10.70$ &$583.410\pm 7.071$ &$649.739\pm 9.875$ &$484.451\pm 5.453$  \\
SDSSJ212005.98$+$115506.4 & $1246.77\pm 10.11$ &$821.366\pm 7.469$ &$789.548\pm 11.20$ &$1341.56\pm 7.007$  \\
SDSSJ230511.16$+$140347.3 & $701.091\pm 7.831$ &$470.729\pm 5.550$ &$437.270\pm 5.566$ &$384.627\pm 5.169$  \\
\hline
\footnotetext[1]{Photometry described in text.}
\footnotetext[2]{The spectrum has no data at $6200\AA<\lambda<7000\AA$.}
\footnotetext[3]{The spectrum of SDSSJ115825.59$+$505501.4 has S/N too low to yield reliable redshift or line measurements.}
\footnotetext[4]{The galaxy appear old and show no emission lines.}
\footnotetext[5]{The image has S/N too low to yield reasonable photometric measurement.}
\end{longtable}
\end{center}
\renewcommand{\thefootnote}{\arabic{footnote}}
\normalsize
\clearpage

\appendix
\section{Matched IRAC and SDSS Photometry Model}
\label{appa}

\par{Photometry of the galaxies studied in this work was measured in the IRAC and SDSS imaging with a exponential ellipsoidal model as described in this Appendix. Since the images of SDSS and IRAC we use in our calculation have similar seeing, the model is not convolved under this approximation. The ellipsoid of the model is centered on each galaxy and has the following form.}
\begin{eqnarray}
f_{m}(x,y) &=& f_{0}e^{-1.68\rho(x,y)}\label{flux}\\
\rho(x,y) &=& [(x\hspace{3mm} y)\cdot{\bf A}\cdot{x \choose y}]^{\frac{1}{2}}\label{r}
\end{eqnarray}
\par{In the above equations, $f_{m}(x,y)$ gives the value of the model flux on each pixel and {\bf A} is a $2\times 2$ tensor with two eigenvectors along the major and minor axes of the ellipse with a position angle $\theta$. The two corresponding eigenvalues are $\frac{1}{a^{2}}$ and $\frac{1}{b^{2}}$ where $a$ and $b$ are the lengths of the major and minor axes as measured in the $r$ band. We adopt $a$, $a/b$ and $\theta$ from the SDSS photometric pipeline~\citep{lupton-aj-1999}.}
\par{$f_{0}$ is the best fitted central pixel flux to each individual image. All magnitudes are calculated in the AB magnitude system. The best value of $f_{0}$ is found with the minimization of the function $\eta^{2}$, an imitation function of $\chi^{2}$ without the usage of the error weighting. The reason why the error weighting does not come into $\eta^{2}$ is to avoid using the to-be-measured quantity in the fitting process, since the errors are going to be measured with the best fit parameter. $\eta^{2}$ has the following form:}
\begin{equation}
\eta^{2}=\sum_{i,j}(f_{m}(x_{i},y_{j})-f_{obs}(x_{i},y_{j}))^{2}\label{eta}
\end{equation}
\par{In eq.(\ref{eta}), $x_{i}$ and $y_{j}$ indicate the location of the pixels. $f_{obs}(x_{i},y_{j})$ gives the observed flux on the corresponding pixel. The best value of $f_{0}$ in eq.(\ref{eta}) minimizes $\eta^{2}$. In the other words, the derivative $\frac{d\eta^{2}}{df_{0}}$ is zero at the best value of $f_{0}$, so that:}
\begin{equation}
f_{0}=\frac{\sum_{i,j}f_{obs}(x_{i},y_{j})e^{-1.68\cdot \rho(x_{i},y_{j})}}{\sum_{k,l}(e^{-1.68\cdot \rho(x_{k},y_{l})})^{2}}.\label{f0}
\end{equation}
\par{This fitting is conducted within the range of $r\le 4$ on the image. With $f_{0}$ given in eq.(\ref{f0}), we can assign a weighting function to each pixel}
\begin{equation}
w_{ij}=e^{1.68\cdot \rho(x_{i},y_{j})}\cdot \frac{\sum_{k,l}e^{-1.68\cdot \rho(x_{k},y_{l})}}{\sum_{k,l}(e^{-1.68\cdot \rho(x_{k},y_{l})})^{2}}.\label{weighting}
\end{equation}
\par{The measured total flux, $F_{est}$ is thus the weighted sum of $f_{obs}$ over the SDSS-smosaic image:}
\begin{equation}
F_{est}=\sum_{i,j}w_{ij}f_{obs}(x_{i},y_{j})\label{fest}
\end{equation}


\begin{thebibliography}{100}
\addcontentsline{toc}{chapter}{Bibliography}

\bibitem[{Abazajian {et~al.}(2004)}]{abazajian-aj-2004} Abazajian, K. {et~al.}, 2004, \aj, 128, 502

\bibitem[{Adelmen-McCarthy {et~al.}(2006)}]{adelmen-mccarthy-apjs-2006} Adelmen--McCarthy,~J.~K. {et~al.}, 2006, \apjs, 162, 38

\bibitem[{Adelmen-McCarthy {et~al.}(2007)}]{adelmen-mccarthy-apjs-2007} Adelmen--McCarthy,~J.~K. {et~al.}, 2007, \apjs, 172, 634

\bibitem[{Allenmandola {et~al.}(1989)}]{allenmandola-apjs-1989} Allenmandola,~L.~J. {et~al.}, 1989, \apjs, 290, L25

\bibitem[{Baldwin {et~al.}(1981)}]{baldwin-pasp-1981} Baldwin,~J. {et~al.}, 1981, \pasp, 93, 5

\bibitem[{Bertin and Arnouts (1996)}]{bertin-aaps-1996} Bertin,~M. and Arnouts,~S., 1996, \aaps, 117, 393

\bibitem[{Blanton {et~al.}(2003a)}]{blanton-aj-2003} Blanton,~M.~R. {et~al.}, 2003, \aj, 124, 2348

\bibitem[{Blanton {et~al.}(2003b)}]{blanton-apj-2003} Blanton,~M.~R. {et~al.}, 2003, \apj, 592, 819

\bibitem[{Blanton {et~al.}(2005a)}]{blanton-apj-2005a} Blanton,~M.~R. {et~al.}, 2005, \apj, 631, 208

\bibitem[{Boselli {et~al.}(1998)}]{boselli-aa-1998} Boselli,~A. {et~al.}, 1998, \aap, 335, 53

\bibitem[{Bresolin (2007)}]{bresolin-apj-2007} Bresolin,~F., 2007, \apj, 656, 186

\bibitem[{Bruzual and Charlot(2003)}]{bruzual-mnras-2003} Bruzual,~G., Charlot~S., 2003, \mnras, 344, 1000

\bibitem[{Calzetti {et~al.}(2005)}]{calzetti-apj-2005} Calzetti,~D. {et~al.}, 2005, \apj, 633, 871

\bibitem[{Calzetti {et~al.}(2007)}]{calzetti-apj-2007} Calzetti,~D. {et~al.}, 2007, \apj, 666, 870

\bibitem[{Dale {et~al.}(2001)}]{dale-aj-2001} Dale,~D.~A. {et~al.}, 2001, \aj, 122, 1736

\bibitem[{Davoodi {et~al.}(2006)}]{davoodi-mnras-2006} Davoodi,~P. {et~al.}, 2006, \mnras, 371, 1113

\bibitem[{Dopita {et~al.}(2006)}]{dopita-apj-2006} Dopita,~M.~A. {et~al.}, 2006, \apj, 647, 244

\bibitem[{Draine and Li (2007)}]{draine-apj-2007a} Draine,~B.~T. and Li,~A., 2007, \apj, 657, 810

\bibitem[{Draine {et~al.}(2007)}]{draine-apj-2007b} Draine,~B.~T. {et~al.}, 2007, \apj, 663, 866

\bibitem[{Engelbracht {et~al.}(2005)}]{engelbracht-apj-2005} Engelbracht,~C.~W. {et~al.}, 2005, \apj, 628, L29

\bibitem[{Engelbracht {et~al.}(2008)}]{engelbracht-apj-2008} Engelbracht,~C.~W. {et~al.}, 2008, \apj, 678, 804

\bibitem[{F\"{o}rster-Schreiber {et~al.}(2004)}]{forster-aa-2004} F\"{o}rster-Schreiber, N.~M. {et~al.}, 2004, \aap, 419, 501

\bibitem[{Gordon {et~al.}(2008)}]{gordon-apj-2008} Gordon,~K.~D. {et~al.}, 2008, \apj, 682, 336

\bibitem[{Helou {et~al.}(2004)}]{helou-apjs-2004} Helou,~G. {et~al.}, 2004, \apjs, 154, 253

\bibitem[{Hogg {et~al.}(2005)}]{hogg-apj-2005} Hogg,~D.~W. {et~al.}, 2005, \apj, 624, 162

\bibitem[{Houck {et~al.}(2004)}]{houck-apjs-2004} Houck,~J.~R. {et~al.}, 2004, \apjs, 154, 211

\bibitem[{Izotov {et~al.}(1997)}]{izotov-apj-1997} Izotov,~Y.~I. {et~al.}, 1997, \apj, 476, 698

\bibitem[{Jackson {et~al.}(2006)}]{jackson-apj-2006} Jackson,~D.~C. {et~al.}, 2006, \apj, 646, 192

\bibitem[{Kauffmann {et~al.}(2003)}]{kauffmann-mnras-2003} Kauffmann,~G. {et~al.}, 2003, \mnras, 346, 1055

\bibitem[{Kennicutt {et~al.}(1994)}]{kennicutt-apj-1994} Kennicutt,~R.~C.,~Jr. {et~al.}, 1994, \apj, 435, 22

\bibitem[{Kennicutt(1998)}]{kennicutt-araa-1998} Kennicutt,~R.~C.,~Jr., 1998, \araa, 36, 189

\bibitem[{Kennicutt {et~al.}(2003)}]{kennicutt-pasp-2003} Kennicutt,~R.~C.,~Jr. {et~al.}, 2003, \pasp, 115, 928

\bibitem[{Kewley {et~al.}(2001)}]{kewley-apj-2001} Kewley,~L. {et~al.}, 2001, \apj, 556, 121

\bibitem[{Lacy {et~al.}(2005)}]{lacy-apjs-2005} Lacy,~M. {et~al.}, 2005, \apjs, 161, 41

\bibitem[{Lamareille {et~al.}(2006)}]{lamareille-aa-2006} Lamareille,~F. {et~al.}, 2006, \aap, 448, 893

\bibitem[{Latter(1991)}]{latter-apj-1991} Latter,~W.~B., 1991, \apj, 377, 187

\bibitem[{Leger and Puget(1984)}]{leger-apj-1984} Leger,~A. and Puget,~J.~L.,~1984,~\apj,~137,~L5

\bibitem[{Leitherer {et~al.}(1999)}]{leitherer-apjs-1999} Leitherer,~C. {et~al.}, 1999, \apjs, 123, 3

\bibitem[{Li and Draine(2002)}]{li-apj-2002} Li,~A.~and Draine,~B.~T., 2002, \apj, 572, 232

\bibitem[{Lonsdale {et~al.}(2003)}]{lonsdale-pasp-2003} Lonsdale,~C. {et~al.}, 2003, \pasp, 154, 54L

\bibitem[{Lupton {et~al.}(1999)}]{lupton-aj-1999} Lupton,~R.~H., Gunn,~J.~E, and Szalay,~A.~S., 1999, \aj, 118, 1406L

\bibitem[{Mas-Hesse and Kunth(1999)}]{mas-hesse-aa-1999} Mas-Hesse,~J.~M. and Kunth,~D., 1999, \aap, 349, 765

\bibitem[{Madden {et~al.}(2006)}]{madden-aa-2006} Madden,~S.~C. {et~al.}, 2006, \aap, 446, 877

\bibitem[{O'Halloran {et~al.}(2006)}]{ohalloran-apj-2006} O'Halloran,~B. {et~al.}, 2006, \apj, 641, 795

\bibitem[{Pahre {et~al.}(2004)}]{pahre-apjs-2004} Pahre,~M.~A. {et~al.}, 2004, \apjs, 154, 235

\bibitem[{Pettini {et~al.}(2004)}]{pettini-mnras-2004} Pettini,~M. and Pagel,~B.~E.~J., 2004, \mnras, 348, L59

\bibitem[{Peeters {et~al.}(2004)}]{peeters-apj-2004} Peeters,~E., Spoon,~H.~W.~W., and Tielens,~A.~G.~G.~M., 2004, \apj, 613, 986

\bibitem[{Pilyugin and Thuan(2006)}]{pilyugin-apj-2005} Pilyugin,~L.~S. and Thuan,~T.~X., 2005, \apj, 631, 231

\bibitem[{Plante and Sauvage(2002)}]{plante-apj-2002} Plante,~A. and Sauvage,~M., 2002, \apj, 124, 1995

\bibitem[{Reach {et~al.}(2005)}]{reach-pasp-2005} Reach,~W.~T. {et~al.}, 2007, \pasp, 117, 978

\bibitem[{Roche {et~al.}(1991)}]{roche-mnras-1991} Roche,~P.~F. {et~al.}, 1991, \mnras, 248, 606

\bibitem[{Rosenberg {et~al.}(2006)}]{rosenberg-apj-2006} Rosenberg,~J.~L. {et~al.}, 2006, \apj, 636, 742

\bibitem[{Schlegel {et~al.}(1998)}]{schlegel-apj-1998} Schlegel,~D.~J., Douglas,~P.~F., and Davis,~M., 1998, \apj, 500, 525

\bibitem[{Smith {et~al.}(2007)}]{smith-apj-2007} Smith,~J.~D. {et~al.}, 2007, \apj, 656, 770

\bibitem[{Struskie {et~al.}(2006)}]{struskie-aj-2006} Struskie,~M.~F. {et~al.}, 2006, \aj, 131, 1163S

\bibitem[{Thuan {et~al.}(1999)}]{thuan-apj-1999} Thuan,~T.~X., Sauvage,~M and Madden,~S., 1999, \apj, 516, 783

\bibitem[{Tielens {et~al.}(1999)}]{tielens-esasp-1999} Tielens,~A.~G.~G.~M. {et~al.}, 1999, ESA SP-427, 579

\bibitem[{Tremonti {et~al.}(2004)}]{tremonti-apj-2004} Tremonti,~C.~A. {et~al.}, 2004, \apj, 613, 898

\bibitem[{Uchida {et~al.}(1998)}]{uchida-apj-1998} Uchida,~K.~I. {et~al.}, 1998, \apj, 493, L109

\bibitem[{Veilleux {et~al.}(1987)}]{veilleux-apjs-1987} Veilleux,~S. {et~al.}, 1987, \apjs, 63, 295

\bibitem[{Werner {et~al.}(2004)}]{werner-apjs-2004} Werner,~M. {et~al.}, 2004, \apjs, 154, 1

\bibitem[{Willick {et~al.}(1997)}]{willick-apj-1997} Willick,~J.~A. {et~al.}, 1997, \apj, 486, 629

\bibitem[{Wu {et~al.}(2005)}]{wu-apj-2005} Wu,~H. {et~al.}, 2005, \apj, 632, L79

\bibitem[{Wu {et~al.}(2007)}]{wuh-apj-2007} Wu,~H. {et~al.}, 2007, \apj, 668, 87

\bibitem[{Wu {et~al.}(2006)}]{wu-apj-2006} Wu,~Y. {et~al.}, 2006, \apj, 639, 157

\bibitem[{Wu {et~al.}(2007)}]{wuy-apj-2007} Wu,~Y. {et~al.}, 2007, \apj, 

\bibitem[{York {et~al.}(2000)}]{york-aj-2000} York,~D. {et~al.}, 2000, \aj, 120, 1579


\end{thebibliography}
\end{document}